\documentclass[12pt]{iopart}

\usepackage{bbm}
\usepackage{graphicx}
\begin{document}

\title[Microscopic scattering theory for interacting bosons in weak random potentials]{Microscopic scattering theory for interacting bosons in  weak random potentials}

\author{Tobias Geiger, Andreas Buchleitner, and Thomas Wellens}

\address{Physikalisches Institut, Albert-Ludwigs-Universit\"at Freiburg, D-79104 Freiburg, Germany}
\ead{Thomas.Wellens@physik.uni-freiburg.de}
\begin{abstract}
We develop a diagrammatic scattering theory for
interacting bosons in a three-dimensional, weakly disordered potential. 
Based on a microscopic $N$-body scattering theory, we identify the relevant diagrams including elastic and inelastic collision processes that are sufficient to describe diffusive quantum transport. By taking advantage of the statistical properties of the weak disorder potential, we demonstrate how the $N$-body dynamics can be reduced to a nonlinear integral equation of Boltzmann type for the single-particle diffusive flux. 
Our theory reduces to the Gross-Pitaevskii mean field description in the limit where only elastic collisions are taken into account. However, even at weak interaction strength, inelastic collisions lead to energy redistribution between the bosons -- initially prepared all at the same single-particle energy -- and thereby induce thermalization of the single-particle current. In addition, we include also weak localization effects and determine the coherent corrections to the incoherent transport in terms of the coherent backscattering signal. We find that inelastic collisions lead to an enhancement of the backscattered cone in a narrow spectral window for increasing interaction strength.
\end{abstract}

\maketitle

\section{Introduction}

In recent years, increasing interest  has been devoted to the behaviour of ultracold atoms in disordered potentials. Whereas the first experiments \cite{clement05,fort05,schulte05} concentrated on the realization of  Anderson localization \cite{anderson58} in one dimension, this intriguing disorder effect -- which leads to complete suppression of diffusive transport due to destructive interference --  has now also been observed in three dimensions \cite{jendrzejewski11,kondov11}. The 3D case is especially interesting since it exhibits a  transition from extended to localized single-particle eigenstates: In the absence of interactions, particles with low energy  are localized, whereas those with higher energy (in comparison with the strength of the disorder potential) propagate diffusively in the random potential. Also in the latter case  -- on which we concentrate in the present paper --  wave interference effects are relevant, though less pronounced: They lead to weak localization \cite{bergmann84} (i.e.~reduction of the diffusion constant instead of complete suppression of diffusion) and, associated with that, coherent backscattering \cite{kuga84,albada85,wolf85} (i.e. enhancement of backscattering), which has recently been observed also with atomic matter waves \cite{labeyrie12,jendrzejewski12, karpiuk12}. Beyond the scope of \cite{labeyrie12,jendrzejewski12} lies the investigation of the interplay between disorder and interactions, where it is not well understood, especially in the higher-dimensional case, to what extent interaction leads to a loss of coherence, i.e.~to a breakdown of localization effects \cite{shepelyansky94, ivanchenko11}. Most theoretical works, e.g.~\cite{lee90,huang92,giorgini94,bilas06,gaul11}, focus on the regime of -- or close to -- thermal equilibrium and examine, e.g., the effect of disorder on the condensate fraction, superfluid fraction or the sound velocity 
\cite{huang92, giorgini94,gaul11}. In this case, weak interactions can usually be treated perturbatively, e.g., by introducing Bogoliubov quasiparticles \cite{zagrebnov01}.

In contrast, the present paper investigates a stationary scattering setup far from thermal equilibrium. Here, bosonic atoms are continuously emitted from a coherent source (\lq atom laser\rq\ \cite{guerin06,couvert08}) and guided into the random potential until a stationary scattering state is reached. Theoretical studies of this scattering scenario so far either neglect the interparticle interaction \cite{kuhn05,kuhn07}, treat it on the mean-field level \cite{paul05, hartung08,wellens09b,wellens09c}, or apply a Hartree-Fock-Bogoliubov approach \cite{ernst10}, which is appropriate in the case of a large condensate fraction. If all atoms enter the scattering region at fixed initial energy, the Gross-Pitaevski equation obtained within the mean-field approach predicts either a stationary regime with the same final energy for all scattered atoms,
or a non-stationary, time-dependent behavior \cite{paul05,ernst10}. In contrast, according to the microscopic scattering theory developed in the present paper, atoms exchange energy with each other due to mutual collision events, leading to strong depletion of the condensate already for small interactions. As shown in \cite{geiger12}, this finally leads to a stationary state with thermal Maxwell-Boltzmann distribution for those atoms which propagate deeply into the scattering region. 

The present paper is devoted to a detailed presentation of the underlying  bosonic many-particle scattering theory. Starting from the $N$-particle Hamiltonian, we derive a nonlinear transport equation for the average particle density. Since this transport equation amounts to a stationary version of the  Boltzmann equation \cite{uehling33}, our approach is, in this respect, comparable to previous works on quantum kinetic equations \cite{spohn07,benedetto08,kirkpatrick83,zaremba99,gardiner97,walser99,schelle11,proukakis01,wachter01}. In contrast to these works, however, 
the additional presence of a disorder potential (apart from the atom-atom interactions) in our setup allows us to quantify the regime of validity of the transport equation in a more rigorous way. In the regime of weak disorder, the disorder average enables us to neglect  correlations between atoms induced by collisions, which in turn is the basic assumption required for reducing a many-particle problem to an effective single-particle description.
Moreover -- and again in contrast to the above works -- we go beyond the case of purely diffusive transport, and also incorporate quantum interference corrections leading to coherent backscattering into our theory.

 Correspondingly, the paper is structured as follows: In Sec.~\ref{sec:single}, we set the stage by reviewing some important aspects of standard scattering theory for a single particle. The case of many interacting particles will be addressed in
Sec.~\ref{sec:many}: Starting from the Hamiltonian including pairwise atom-atom interaction, we introduce a diagrammatic notation for the transition amplitudes of many particles, from which the scattered flux density can be calculated after taking the trace over the undetected particles. As we will see, this trace leads to a distinction of atom-atom collisions events into inelastic and elastic collisions, respectively, where the latter are shown to reproduce the mean-field description given by the Gross-Pitaevskii equation. Whereas the methods presented up to Sec.~\ref{sec:many} are generally valid for an arbitrary scattering potential, we focus on the case of a weak random potential from Sec.~\ref{sec:incoherent} on. The assumption of weak disorder   ($k \ell_{\rm dis}\gg 1$ with wavenumber $k$ and disorder scattering mean free path $\ell_{\rm dis}$) is crucial, since it allows to reduce the --  in principle infinitely complicated \cite{erdoes12} -- hierarchy of many-particle diagrams to a tractable subclass of diagrams, i.e.~ladder and crossed diagrams \cite{akkermans07}, which are composed out of a small number of building blocks. As shown in Sec.~\ref{sec:incoherent}, the sum of all ladder diagrams amounts to a Boltzmann-like equation for diffusive transport eventually leading to complete thermalization due to inelastic atom-atom collisions in case of an infinitely large scattering region. Sec.~\ref{sec:coherent} is devoted to the derivation of transport equations describing coherent backscattering based on crossed diagrams. Finally, in Sec.~\ref{sec:results} we present the results of numerical solutions of the ladder and crossed transport equations exemplifying the behaviour of diffusive transport for a finitely large scattering region, and the effect of elastic and inelastic atom-atom collisions on coherent backscattering, respectively. Sec.~\ref{sec:concl} concludes the paper. Several technical aspects are relegated to Appendices A-E.

\section{Scattering theory for a single particle}
\label{sec:single}

We write the Hamiltonian for a single particle in the following form:
\begin{equation}
\hat{H}=\hat{H}_0+\hat{V}\label{eq:hamiltoniansingle}\,,
\end{equation}
where $\hat{H}_0$ denotes free propagation and $\hat{V}$ the disorder potential. The eigenstates $|{\bf k}\rangle$ of $\hat{H}_0$ are plane waves with wave vector $\bf k$:
\begin{equation}
\hat{H}_0=\int\frac{{\rm d}{\bf k}}{(2\pi)^3} E_{\bf k}|{\bf k}\rangle\langle {\bf k}|\label{eq:h0single}\,,
\end{equation}
and energy
\begin{equation}
E_{\bf k}=k^2\label{eq:energy}\,,
\end{equation}
where we set $\hbar^2/(2m)\equiv 1$. The matrix elements of $\hat{V}$ are given by the Fourier transform of the disorder potential $V({\bf r})$:
\begin{equation}
\langle{\bf k}_2|\hat{V}|{\bf k}_1\rangle=\int {\rm d}{\bf r}\,V({\bf r})e^{i({\bf k}_1-{\bf k}_2){\bf r}}
\label{eq:vsingle} \,.
\end{equation}
In order to obtain a properly defined scattering scenario, we assume that $V({\bf r})$ is non-zero only inside a finite scattering region ${\mathcal V}$. This allows us to define an asymptotically free initial state:
\begin{equation}
|i_1\rangle=\int \frac{{\rm d}{\bf k}}{(2\pi)^{3}}~w({\bf k})
|{\bf k}\rangle\label{eq:initial1}\,,
\end{equation}
with normalized wavepacket $w({\bf k})$, i.e. 
$\int{\rm d}{\bf k} |w({\bf k})|^2 =(2\pi)^3$, which we assume to be a quasi-monochromatic wavepacket, i.e., sharply peaked around the initial wavevector ${\bf k}_i$ with energy $E_i=k_i^2$, see Eq.~(\ref{eq:energy}). Therefore, the spatial density resulting from the Fourier transform of $w({\bf k})$:
\begin{equation}
|\widetilde{w}({\bf r})|^2=\left|\int\frac{{\rm d}{\bf k}}{(2\pi)^3}e^{i{\bf k}\cdot{\bf r}}w({\bf k})\right|^2\simeq \left|\int\frac{{\rm d}{\bf k}}{(2\pi)^3}w({\bf k})\right|^2\label{eq:density1}
\end{equation}
is approximately constant inside the scattering region, i.e. for ${\bf r}\in{\mathcal V}$. If the state $\exp(-i\hat{H}_0T)|i_1\rangle$ is prepared at time $T\to-\infty$, the wavepacket arrives at the scattering region at time $t=0$, and a quasi-stationary scattering state
\begin{equation}
|f_{+,1}\rangle=\hat{\Omega}_+^{(V)}(E_i)|i_1\rangle\label{eq:fplus1}
\end{equation}
is reached at that time. Here, the operator $\hat{\Omega}_+^{(V)}(E)$ is defined by
\begin{equation}
\hat{\Omega}_+^{(V)}(E)={\mathbbm 1}+\hat{G}_V(E)\hat{V}\label{eq:omegav}\,,
\end{equation}
where 
\begin{equation}
\hat{G}_V(E)=\frac{1}{E-\hat{H}_0-\hat{V}+i\epsilon}\label{eq:gv}\,,
\end{equation}
with infinitesimally small $\epsilon>0$,
denotes the (retarded) Green's operator associated to the Hamiltonian $\hat{H}=\hat{H}_0+\hat{V}$. 
The operators $\hat{\Omega}_+^{(V)}(E)$ and $\hat{G}_V(E)$ fulfill the following versions of the Lippmann-Schwinger equation:
\begin{eqnarray}
\hat{\Omega}_+^{(V)}(E) & = & {\mathbbm 1}+\hat{G}_0(E)\hat{V}\hat{\Omega}_+^{(V)}(E)\label{eq:lippmannschwingerv}\,,\\
\hat{G}_V(E) & = & \hat{G}_0(E)+\hat{G}_0(E)\hat{V}\hat{G}_V(E)\label{eq:lippmannschwingergv}\,,
\end{eqnarray}
where $\hat{G}_0(E)$ denotes the vacuum Green's operator:
\begin{equation}
\hat{G}_0(E)=\frac{1}{E-\hat{H}_0+i\epsilon}\label{eq:g0}\,.
\end{equation}
The operator $\hat{\Omega}_+^{(V)}(E)$ is closely related
to the M\o ller operator $\hat{\Omega}_+^{(V)}=\lim_{T\to-\infty}\exp[i (\hat{H}_0+\hat{V})T)\exp(-i\hat{H}_0T)$, as their action on an eigenstate $|\psi\rangle$ of $\hat{H}_0$ with energy $E$ is identical, i.e. $\hat{\Omega}_+^{(V)}|\psi\rangle=\hat{\Omega}_+^{(V)}(E)|\psi\rangle$ if $\hat{H}_0|\psi\rangle=E|\psi\rangle$.
Since, in the following, we will apply $\hat{\Omega}_+^{(V)}(E)$ only to such eigenstates -- or quasi-eigenstates, as $|i_1\rangle$ in Eq.~(\ref{eq:fplus1}) -- we will henceforth  refer also to $\hat{\Omega}_+^{(V)}(E)$ as \lq M\o ller operator\rq. Finally,  the expectation value of an arbitrary observable $\hat{A}$ in the (quasi-)stationary scattering state results as $\langle \hat{A}\rangle=\langle f_{+,1}|\hat{A}|f_{+,1}\rangle$.

Let us note that, instead of using the M\o ller operator, a scattering process can also be characterized by the $S$-matrix, 
$\hat{S}=\left(\hat{\Omega}_-^{(V)}\right)^\dagger \hat{\Omega}_+^{(V)}$ (where $\hat{\Omega}_-^{(V)}$ is defined in the same way as $\hat{\Omega}_+^{(V)}$, but with $T\to+\infty$ instead of $-\infty$). We could formulate the following $N$-particle scattering theory equally well in terms of the $S$-matrix. However, since the $S$-matrix maps incoming onto outgoing asymptotically free states, it does not allow -- in contrast to the M\o ller operator -- to evaluate what is happening {\em inside} the scattering region, e.g. to calculate the (quasi-)stationary density or flux of particles inside $\mathcal V$. For this reason, we prefer using the (quasi-)stationary scattering state $|f_{+,1}\rangle$, see Eq.~(\ref{eq:fplus1}) (or its $N$-particle counterpart $|f_+\rangle$, see Eq.~(\ref{eq:fplus}) below) in the following.

\section{Scattering theory for many bosonic particles}
\label{sec:many}

\subsection{Many-particle Hamiltonian}
\label{sec:manyham}

We
add a term $\hat{U}$ to the Hamiltonian, Eq.~(\ref{eq:hamiltoniansingle}), denoting the interaction between particles:
\begin{equation}
\hat{H}   = 
\hat{H}_0 + \hat{V} +\hat{U}\label{eq:hamiltonian}\,.
\end{equation}
As compared to Eqs.~(\ref{eq:h0single},\ref{eq:vsingle}), the operators $\hat{H}_0$ and $\hat{V}$ are
generalized as follows to the many-particle Hilbert space:
\begin{eqnarray}
\hat{H}_0 & = & \int\frac{{\rm d}{\bf k}}{(2\pi)^3} E_{\bf k}\hat{a}_{{\bf k}}^\dagger \hat{a}_{{\bf k}}\label{eq:h0}\,,\\
\hat{V} & = & \int {\rm d}{\bf r}~V({\bf r}) \hat{\psi}^\dagger({\bf r})\hat{\psi}({\bf r})\label{eq:v}\,,
\end{eqnarray}
with creation and annihilation operators $\hat{a}^\dagger_{\bf k}$ and $\hat{a}_{\bf k}$ for particles with wave vector ${\bf k}$,
whereas the operators $\hat{\psi}({\bf r})=\int{\rm d}{\bf k}\exp(i{\bf k}\cdot{\bf r})\hat{a}_{\bf k}/(2\pi)^3$ 
and $\hat{\psi}^\dagger({\bf r})=\int{\rm d}{\bf k}\exp(-i{\bf k}\cdot{\bf r})\hat{a}_{\bf k}^\dagger/(2\pi)^3$
annihilate and create, respectively, a particle at position ${\bf r}$. 

In contrast to $\hat{H}_0$ and $\hat{V}$, the interaction $\hat{U}$ acts on two particles:
\begin{equation}
\hat{U}=\frac{1}{2}\int{\rm d}{\bf r}_1{\rm d}{\bf r}_2~U({\bf r}_1-{\bf r}_2) \hat{\psi}^\dagger({\bf r}_1)\hat{\psi}^\dagger({\bf r}_2)
\hat{\psi}({\bf r}_2)\hat{\psi}({\bf r}_1)\label{eq:u}\,,
\end{equation}
with atom-atom interaction potential $U({\bf r})$.
In the following, a collision event between two particles will be described by the  
$T$-matrix \cite{taylor}:
\begin{equation}
\hat{T}_U(E)=\hat{U}+\hat{U}\hat{G}_0(E)\hat{U}+\hat{U}\hat{G}_0(E)\hat{U}\hat{G}_0(E)\hat{U}+\dots\label{eq:tmatrixdef}\,.
\end{equation}
According to Eq.~(\ref{eq:tmatrixdef}), the matrix elements of $\hat{T}_U(E)$ with respect to two-particle states describe repeated application of the interaction $\hat{U}$ on the same pair of particles, interrupted by free propagation $\hat{G}_0(E)$. Separating the center-of-mass from the relative coordinates, the two-body $T$ matrix fulfills momentum conservation:
\begin{equation}
\langle{\bf k}_3,{\bf k}_4|\hat{T}_U(E)|{\bf k}_1, {\bf k}_2\rangle=
(2\pi)^3 \delta({\bf k}_1+{\bf k}_2-{\bf k}_3-{\bf k}_4) 
\langle {\bf k}_{34}|\hat{T}^{(1)}_U(E_{12})|{\bf k}_{12}\rangle\label{eq:tmatrix}\,,
\end{equation}
where $\hat{T}^{(1)}_U(E_{12})$ is the $T$-matrix for a single particle (with reduced mass $m/2$) scattered by the potential $U({\bf r})$
at energy $E_{12}=E-({\bf k}_1+{\bf k}_2)^2/2$,
$|{\bf k}_{12}\rangle  =  \Bigl(\left|({\bf k}_1-{\bf k}_2)/2\right>+\left| ({\bf k}_2-{\bf k}_1)/2\right>\Bigr)/\sqrt{2}$, and
$|{\bf k}_{34}\rangle = \Bigl(\left|({\bf k}_3-{\bf k}_4)/2\right>+\left| ({\bf k}_4-{\bf k}_3)/2\right>\Bigr)/\sqrt{2}$.
The single-particle $T$-matrix, in turn, fulfills the optical theorem \cite{taylor}:
\begin{equation}
\left(\hat{T}^{(1)}_U(E)\right)^\dagger\left(\hat{G}_{0,m/2}^\dagger(E)-\hat{G}_{0,m/2}(E)\right)\hat{T}^{(1)}_U(E)=\left(\hat{T}^{(1)}_U(E)\right)^\dagger-\hat{T}^{(1)}_U(E)\label{eq:opttheorem}\,,
\end{equation}
expressing conservation of the particle and the energy flux (where $\hat{G}_{0,m/2}$ denotes the vacuum Green's operator for a particle with mass $m/2$ and corresponding dispersion relation $E=2k^2$).

Our many-particle scattering theory presented below, and in particular the transport equations in Secs.~\ref{sec:incoherent} and \ref{sec:coherent}, are valid for an arbitrary interaction potential $U({\bf r})$ -- as long as it is sufficiently weak in the sense specified below (mean distance between collision events larger than between disorder scattering events). Only for the numerical results presented in Sec.~\ref{sec:results}, we will assume a short-range potential with corresponding $s$-wave scattering approximation, see Eq.~(\ref{eq:swave}).

Finally, we note that, in principle, the vacuum $T$-matrix as defined in Eq.~(\ref{eq:tmatrixdef}) is modified by the presence of the disorder potential. To take this into account, the vacuum Green's operator $\hat{G}_0(E)$ must be replaced by the disorder Green's operator $\hat{G}_V(E)$, see Eq.~(\ref{eq:gv}), in Eq.~(\ref{eq:tmatrixdef}). However, since the present paper assumes the case of a very weak disorder potential, 
we will neglect the disorder during each collision event in the following, and therefore use the vacuum $T$-matrix as introduced above.
This approximation is valid if the range of the interaction potential $U({\bf r})$ is much smaller than the disorder mean free path $\ell_{\rm dis}$ introduced in Sec.~\ref{sec:incoherent}.

\subsection{Many-particle transition amplitudes}
\label{sec:nparticle}

We now generalize the scattering scenario outlined in Sec.~\ref{sec:single} to the case of many particles. For this purpose, we assume that, both, the disorder and the particle-particle interaction are non-zero only inside a finite region ${\mathcal V}$ (which, for simplicity, we assume to be the same for $\hat{V}$ and $\hat{U}$). Note that the introduction of a finite interaction region in principle breaks translational invariance, and therefore the $\delta$-function expressing momentum conservation in Eq.~(\ref{eq:tmatrix}) turns into an approximate $\delta$-function. Since, however, we assume the size $L$ of the scattering region $\mathcal V$ to be much larger than the disorder mean free path, i.e. $L\gg \ell_{\rm dis}\gg k^{-1}$ (see below), we can safely neglect the associated small width ($\propto 1/L$) of this $\delta$-function, and still work with the $T$-matrix as given by Eq.~(\ref{eq:tmatrix}).

Our  initial state for $N$ particles reads:
\begin{equation}
|i\rangle=\frac{1}{\sqrt{N!}}\int \frac{{\rm d}{\bf k}_1\dots {\rm d}{\bf k}_N}{(2\pi)^{3N}}~w({\bf k}_1)\dots w({\bf k}_N)
|{\bf k}_1,\dots,{\bf k}_N\rangle\label{eq:initial}\,,
\end{equation}
where all $N$ atoms are described by the same quasi-monochromatic single-atom wavepacket $w({\bf k})$ as given in Eq.~(\ref{eq:initial1}). The factor $1/\sqrt{N!}$ arises from the indistinguishability of bosonic particles.
The corresponding density of particles reads:
\begin{equation}
\rho_0=\langle i|\hat{\psi}^\dagger({\bf r})\hat{\psi}({\bf r})|i\rangle\simeq N\left|\int\frac{{\rm d}{\bf k}}{(2\pi)^3}w({\bf k})\right|^2\label{eq:rho0}\,.
\end{equation}
As mentioned above, this density is approximately uniform within the whole
scattering region ${\mathcal V}$ for a wavepacket sharply peaked around the initial wavevector ${\bf k}_i$. 
Since, in this quasi-monochromatic limit, the density, Eq.~(\ref{eq:density1}), for $N=1$ approaches zero (since the wave packet is spread over an increasingly large region of space), the number $N$ of particles correspondingly must tend to infinity in order to obtain a finite density $\rho_0$.

The M\o ller operator, which yields the 
 quasi-stationary $N$-particle scattering state
\begin{equation}
|f_+\rangle=\hat{\Omega}_+(N E_i)|i\rangle\label{eq:fplus}\,,
\end{equation}
is defined in the same way as above, see Eqs.~(\ref{eq:omegav},\ref{eq:gv}) but  with $\hat{V}+\hat{U}$ instead of $\hat{V}$.
 It therefore fulfills  the Lippmann-Schwinger equation: 
\begin{equation}
\hat{\Omega}_+(E)={\mathbbm 1}+\hat{G}_0(E)\left(\hat{V}+\hat{U}\right)\hat{\Omega}_+(E)\label{eq:lippmannschwinger0}\,,
\end{equation}
which, using Eqs.~(\ref{eq:omegav},\ref{eq:lippmannschwingergv}), can be rewritten as:
\begin{equation}
\hat{\Omega}_+(E)=\hat{\Omega}_+^{(V)}(E)+\hat{G}_V(E)\hat{U}\hat{\Omega}_+(E)\label{eq:lippmannschwinger}\,.
\end{equation}
Iteration of  Eq.~(\ref{eq:lippmannschwinger}) yields an expansion in powers of $\hat{U}$:
\begin{equation}
\hat{\Omega}_+(E) = \hat{\Omega}_+^{(V)}(E)+\hat{G}_V(E)\hat{U} \hat{\Omega}_+^{(V)}(E)+
\hat{G}_V(E)\hat{U} \hat{G}_V(E)\hat{U} \hat{\Omega}_+^{(V)}(E)+\,\dots\label{eq:gseries}\,.
\end{equation}
Remember that, according to Eq.~(\ref{eq:u}), each operator $\hat{U}$ annihilates and creates two particles.
In contrast, the Green's operator $\hat{G}_V$ and the M\o ller operator $\hat{\Omega}_+^{(V)}$ act on all $N$ particles. However, since
these operators describe non-interacting particles, they can be factorized into single-particle operators. As an example, we give here the factorization formulas for the case $N=2$:
\begin{eqnarray}
&&\langle {\bf k}_3,{\bf k}_4|\hat{\Omega}_+^{(V)}(E_{{\bf k}_1}+E_{{\bf k}_2})|{\bf k}_1,{\bf k}_2\rangle  =  \langle {\bf k}_3|\hat{\Omega}_+^{(V)}(E_{{\bf k}_1})|{\bf k}_1\rangle\langle {\bf k}_4|\hat{\Omega}_+^{(V)}(E_{{\bf k}_2})|{\bf k}_2\rangle\nonumber\\
&&  \ \ \ \ \ \ \ \  \ \ \ \ \ \ \ \  \ \ \ \ \ \ \ \  \ \ \ \ \ \ \ \ +  \langle {\bf k}_4|\hat{\Omega}_+^{(V)}(E_{{\bf k}_1})|{\bf k}_1\rangle\langle {\bf k}_3|\hat{\Omega}_+^{(V)}(E_{{\bf k}_2})|{\bf k}_2\rangle
\label{eq:factorizeom}\,,
\end{eqnarray}
and 
\begin{eqnarray}
& & \langle {\bf k}_3,{\bf k}_4|\hat{G}_V(E)|{\bf k}_1,{\bf k}_2\rangle  =  
\frac{1}{(-2\pi i)}\int_{-\infty}^\infty {\rm d}E'~\Bigl[\langle {\bf k}_3|\hat{G}_V(E')|{\bf k}_1\rangle\Bigr.\nonumber\\
& & \ \ \ \ \ \ \ \ \Bigl.\times\langle {\bf k}_4|\hat{G}_V(E-E')|{\bf k}_2\rangle+\langle {\bf k}_4|\hat{G}_V(E')|{\bf k}_1\rangle\langle {\bf k}_3|\hat{G}_V(E-E')|{\bf k}_2\rangle\Bigr]
\label{eq:factorize}\,.
\end{eqnarray}
As mentioned above, the energy argument of our M\o ller operator, Eq.~(\ref{eq:omegav}), is always fixed to the energy of the state it acts on.
In contrast, Green's operators also act on states with different energies. Hence, the energy $E$ of a two-particle Green's operator 
has to be distributed among two one-particle Green's operators  according to Eq.~(\ref{eq:factorize}). 

\begin{figure}
\centerline{\includegraphics[width=4cm]{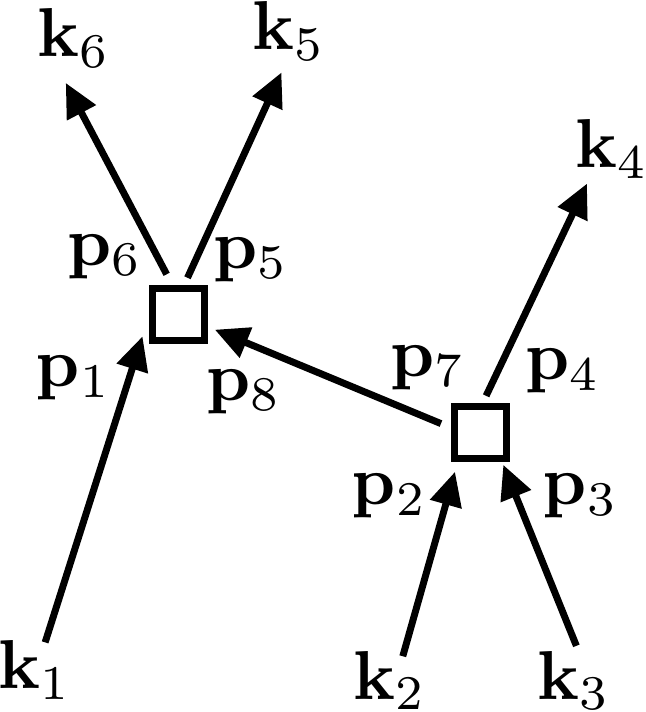}}
\caption{Example of a three-particle scattering process with initial state $|{\bf k}_1,{\bf k}_2,{\bf k}_3\rangle$ and final state
$|{\bf k}_4,{\bf k}_5,{\bf k}_6\rangle$.
 The three arrows associated with the initial state 
represent the M\o ller operator $\hat{\Omega}_+^{(V)}(E_i)$ of the disorder potential, see Eq.~(\ref{eq:omegav}), whereas the remaining arrows refer to the disorder Green's operator $\hat{G}_V$, Eq.~(\ref{eq:gv}). Squares correspond to the two-body $T$-matrix of the particle-particle interaction, Eq.~(\ref{eq:tmatrix}). The transition amplitude corresponding to this scattering process is given in Eq.~(\ref{eq:smatrix3}).}
\label{fig:amplitudes}
\end{figure}

Using the above factorization formulas -- and analogous ones for $N>2$ (see \ref{sec:factorization}) -- we obtain well-defined paths for individual particles between the two-particle interaction events $\hat{U}$. Repeated interaction between the same pair of particles is included in the $T$-matrix, see Eq.~(\ref{eq:tmatrixdef}) (and the discussion at the end of Sec.~\ref{sec:manyham}). We hence replace two-particle matrix elements of $\hat{U}$ by matrix elements of $\hat{T}_U(E)$ (with appropriately defined two-particle energy $E$, see below) in Eq.~(\ref{eq:gseries}), and thereby obtain a sequence of collision events between different pairs of particles. An example of a three-particle scattering process 
is demonstrated in Fig.~\ref{fig:amplitudes}. 
As shown in \ref{sec:factorization}, this diagram gives rise to the following contribution to the transition amplitude:
\begin{eqnarray}
& & \hspace{-10mm}\langle{\bf k}_4,{\bf k}_5,{\bf k}_6|\hat{\Omega}_+^{\rm (fig.\ref{fig:amplitudes})}(3E_i)|{\bf k}_1,{\bf k}_2,{\bf k}_3\rangle   =  
\int_{-\infty}^\infty \frac{{\rm d}E_4{\rm d}E_5}{(-2\pi i)^2}\int\frac{{\rm d}{\bf p}_1\dots {\rm d}{\bf p}_8}{(2\pi)^{24}}
\nonumber\\
& & \hspace{-5mm}\times \langle{\bf k}_6|\hat{G}_V(3E_i-E_4-E_5)|{\bf p}_6\rangle\langle{\bf k}_5|\hat{G}_V(E_5)|{\bf p}_5\rangle
\langle{\bf k}_4|\hat{G}_V(E_4)|{\bf p}_4\rangle\nonumber\\
& & \hspace{-5mm}\times\langle{\bf p}_5,{\bf p}_6|\hat{T}_U(3E_i-E_4)|{\bf p}_1,{\bf p}_8\rangle\langle{\bf p}_8|\hat{G}_V(2 E_i-E_4)|{\bf p}_7\rangle
\langle{\bf p}_4,{\bf p}_7|\hat{T}_U(2E_i)|{\bf p}_2,{\bf p}_3\rangle\nonumber\\
& & \hspace{-5mm}\times\langle{\bf p}_1|\hat{\Omega}_+^{(V)}(E_i)|{\bf k}_1\rangle
\langle{\bf p}_2|\hat{\Omega}_+^{(V)}(E_i)|{\bf k}_2\rangle\langle{\bf p}_3|\hat{\Omega}_+^{(V)}(E_i)|{\bf k}_3\rangle
\label{eq:smatrix3}\,,
\end{eqnarray}
with $E_{{\bf k}_1}\simeq E_{{\bf k}_2}\simeq E_{{\bf k}_3}\simeq E_i$, according to our above assumption of a quasi-monochromatic wavepacket.

In general, the rules for constructing an arbitrary $N$-particle scattering amplitude for a given diagram are as follows:
(i) Apply the disorder M\o ller operator $\hat{\Omega}_+^{(V)}(E_i)$, see Eq.~(\ref{eq:omegav}), to each initial single-particle state $|{\bf k}_1\rangle,\dots,|{\bf k}_N\rangle$. 
The energy associated to each initial particle is given by $E_i$.
(ii) Integrate over all intermediate particles (${\bf p}_1,\dots,{\bf p}_8$ in Fig.~\ref{fig:amplitudes}). 
(iii) Write down the corresponding two-body $T$-matrix element, see Eq.~(\ref{eq:tmatrix}), for any collision between two particles. The energy argument of $\hat{T}_U$ is given by the sum of the two incoming single-particle energies. (iv) 
For each $\hat{T}_U(E)$, write down an integral $\int_{-\infty}^\infty {\rm d}E'/(-2\pi i)$ which determines the energy arguments of the Green's operators 
$\hat{G}_V(E')$ and $\hat{G}_V(E-E')$, see Eq.~(\ref{eq:factorize}), for the two particles after the collision. (v) These two particles may then collide with other particles, and so on ...~.

The total transition amplitude defining the stationary scattering state $|f_+\rangle$, see Eq.~(\ref{eq:fplus}), is then obtained by summing the contributions from all possible different diagrams. For example, in addition to the diagram shown in Fig.~\ref{fig:amplitudes}, eight more diagrams obtained by exchanging the initial and/or final wavevectors $({\bf k}_1,{\bf k}_2,{\bf k}_3)$ and  $({\bf k}_4,{\bf k}_5,{\bf k}_6)$ also contribute to $|f_+\rangle$. 

\subsection{Scattered flux}

As the finally measured quantity, we determine the expectation value of the flux density operator
\begin{equation}
\hat{\bf J}({\bf r}) = 2{\rm Im}\left(\hat{\psi}^\dagger({\bf r})\nabla\hat{\psi}({\bf r})\right)=
\int\frac{{\rm d}{\bf k}{\rm d}{\bf k}'}{(2\pi)^6}~\left(\frac{{\bf k}+{\bf k}'}{2}\right) e^{-i({\bf k}-{\bf k}')\cdot {\bf r}}
\hat{a}^\dagger_{{\bf k}} \hat{a}_{{\bf k}'}\,,
\end{equation} 
with respect to the stationary scattering state $|f_+\rangle$. Since $\hat{\bf J}({\bf r})$ is a one-particle operator, this
implies a partial trace of the density matrix $|f_+\rangle\langle f_+|$ over
$N-1$ undetected particles:
\begin{eqnarray}
{\bf J}({\bf r}) & = & \langle f_+|\hat{\bf J}({\bf r})|f_+\rangle\nonumber\\
& = & \frac{N}{N!}\int\frac{{\rm d}{\bf k}{\rm d}{\bf k}'}{(2\pi)^6}
\left(\frac{{\bf k}+{\bf k}'}{2}\right) e^{-i({\bf k}-{\bf k}')\cdot {\bf r}}
\nonumber\\
& & \times \int \frac{{\rm d}{\bf k}_1\dots{\rm d}{\bf k}_{N-1}}{(2\pi)^{3(N-1)}} 
\langle {\bf k}_1,\dots,{\bf k}_{N-1},{\bf k}'|f_+\rangle\langle f_+|{\bf k}_1,\dots,{\bf k}_{N-1},{\bf k}\rangle\label{eq:trace}\,.
\end{eqnarray}
Placing the detector at position $\bf R$ in the far field of the scattering region (i.e. $|{\bf R}|\gg |{\bf r}|$ for ${\bf r}\in{\mathcal V})$, the scattered flux is finally expressed as a dimensionless quantity (the so-called \lq bistatic coefficient\rq\ \cite{ishimaru}):
\begin{equation}
\gamma(\hat{\bf k}_d)=\lim_{R\to\infty}\left({\bf R}\cdot {\bf J}({\bf R}) \frac{4\pi R}{{\mathcal A}\rho_0\sqrt{E}_i}\right)\label{eq:gamma}\,,
\end{equation}
normalized with respect to the incident flux ${\mathcal A}\rho_0\sqrt{E_i}$, where ${\mathcal A}$ denotes the transverse area (with respect to the incident wave) of the scattering volume $\mathcal V$, and $\hat{\bf k}_d={\bf R}/|{\bf R}|$ is the direction of the detected particle's wavevector. The limit $R\to\infty$ is to be taken {\em after} the quasi-stationary limit $N\to\infty$, see the discussion after Eq.~(\ref{eq:rho0}). Apart from the total flux density $\gamma(\hat{\bf k}_d)$, we will also be interested in the {\em spectral} density $\gamma_{E}(\hat{\bf k}_d)$, i.e. the flux of particles scattered into direction $\hat{\bf k}_d$ with energy $E$, which is given  by:
\begin{equation}
\gamma_{E}(\hat{\bf k}_d) = \lim_{R\to\infty}
\int\frac{{\rm d}{\bf k}{\rm d}{\bf k}'}{16\pi^5}~({\bf R}\cdot{\bf K}) e^{-i({\bf k}-{\bf k}')\cdot {\bf R}}
\frac{\langle f_+|\hat{a}^\dagger_{{\bf k}} \hat{a}_{{\bf k}'}|f_+\rangle}{{\mathcal A}\rho_0\sqrt{E_i}/R}\delta(E-K^2)\label{eq:gammae}\,,
\end{equation} 
where ${\bf K}=({\bf k}+{\bf k}')/2$,
such that $\int_0^\infty {\rm d}E~\gamma_{E}(\hat{\bf k}_d)=\gamma(\hat{\bf k}_d)$.

The factor $1/N!$ in Eq.~(\ref{eq:trace}) arises from the indistinguishability of the bosonic particles. It turns out, however, that this factor -- 
together with the factors $1/\sqrt{N!}$ in Eq.~(\ref{eq:initial}) -- is exactly counterbalanced once we sum the amplitudes 
of all processes where the initial and/or final particles are exchanged. In total, we get the same result as if the particles were distinguishable. This equivalence is generally valid if all particles are prepared in the same initial state, and if the Hamiltonian is symmetric under exchange of particles \cite{tichy10}.

Remember that the number $N$ of particles tends to infinity in the quasi-stationary limit, whereas, in case of a finite scattering region, only a finite number of particles will eventually interact with the finally detected particle. The evolution of the remaining particles (which do not interact with the detected particle) does not influence the result of the partial trace, Eq.~(\ref{eq:trace}).
This follows from the factorization property, Eq.~(\ref{eq:factorizeom}), and the left-unitarity, $\left(\hat{\Omega}_+\right)^\dagger \hat{\Omega}_+={\mathbbm 1}$ of the M\o ller operator. Consequently, in order to calculate the detection signal, we may disregard all scattering processes concerning those particles which do not interact (neither in $|f_+\rangle$ nor in $\langle f_+|$) with the detected particle. (The presence of these particles only leads to a prefactor giving rise to the correct dependence of a given scattering diagram on the density $\rho_0$, see the discussion at the end of \ref{sec:examplediag}.)

\subsection{Trace over undetected particles}

\begin{figure}
\centerline{\includegraphics[width=12cm]{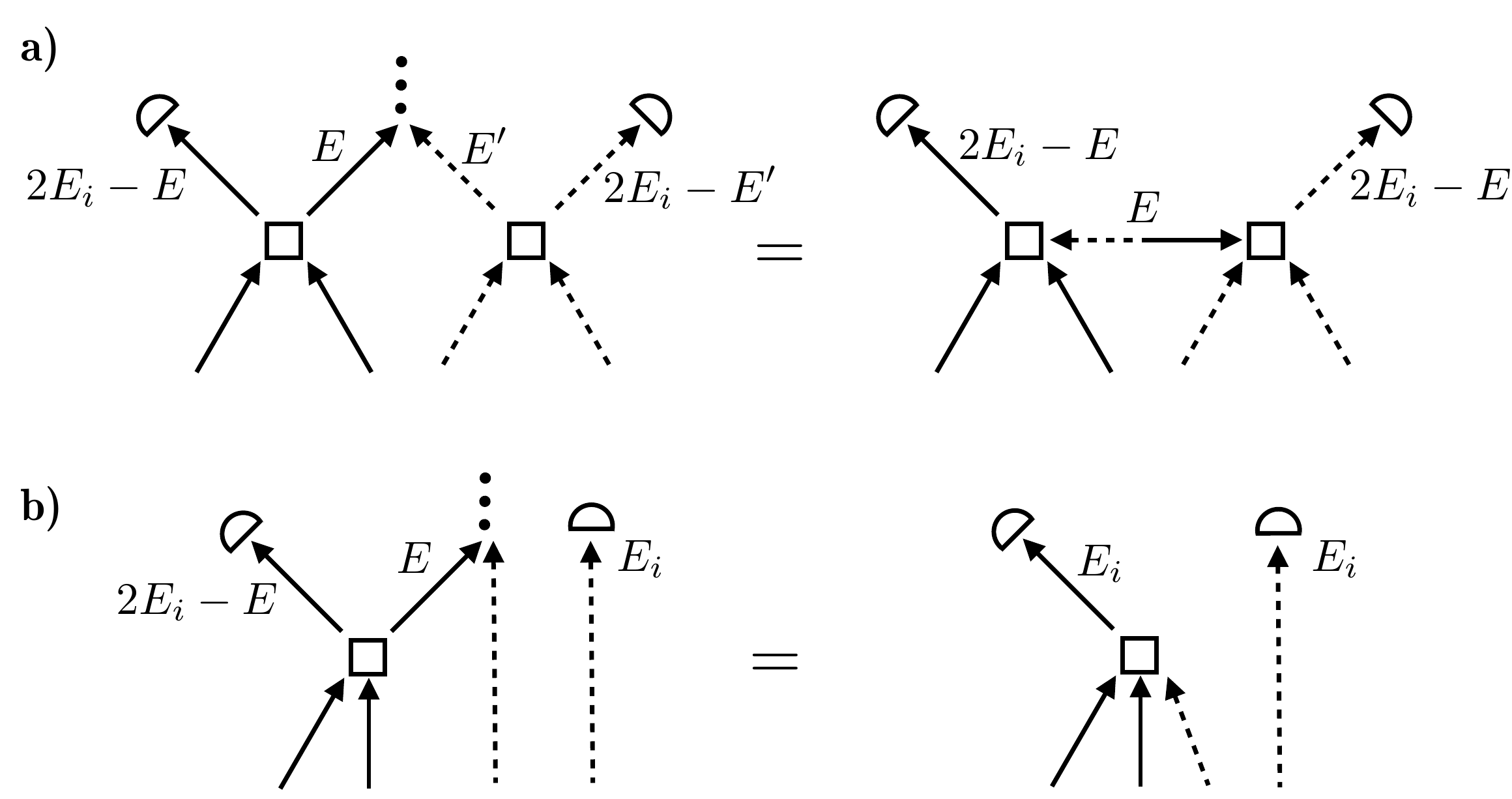}}
\caption{Graphical equations exemplifying the trace over the undetected particles. Arrows and squares refer to single-particle propagators and two-body $T$-matrices, as defined in Fig.~\ref{fig:amplitudes}. Dashed arrows correspond to adjoint propagators $\hat{G}_V^\dagger$
and $\left(\hat{\Omega}_+^{(V)}\right)^\dagger$.
The half circle symbol denotes the detector, whereas the dots on the left-hand side of both equations represent the trace over the undetected particle. {\bf a)} Inelastic scattering of two particles. On the right-hand side, the trace has been performed using Eq.~(\ref{eq:trace3}). This results in the dashed-solid double arrow representing the spectral function $\left[\hat{G}_V^\dagger(E)-\hat{G}_V(E)\right]/(2\pi i)$. The energy of the detected particle is $2E_i-E$.
{\bf b)} Elastic scattering of two particles. The trace is performed using Eq.~(\ref{eq:trace4}). The resulting diagram on the right-hand side is equivalent to a diagram obtained from the Gross-Pitaevski equation. Since, in contrast to a), the conjugate particles (dashed lines) do not undergo a collision, the energy of the detected particle is unchanged ($E_i$).}
\label{fig:trace1}
\end{figure}

According to the recipe given above, the flux density for an arbitrary $N$-particle scattering process is obtained as follows: take a diagram contributing to $|f_+\rangle$, a conjugate diagram
contributing to $\langle f_+|$, apply the observable $\hat{\bf J}({\bf r})$ to one of the final particles of both diagrams, and trace over the undetected particles. An example for two particles is shown in Fig.~\ref{fig:trace1}a) (left-hand side). 
Since both conjugate diagrams (solid and dashed lines, respectively) exhibit a collision event, which redistributes the energy among the two particles according to the factorization formula, Eq.~(\ref{eq:factorize}), the energy of the detected particle is different from the initial energy $E_i$. For this reason, we call this scattering process \lq inelastic\rq. This means that the energies of the single particles change -- although their sum remains conserved. In contrast, Fig.~\ref{fig:trace1}b) shows an elastic scattering process. Here, the conjugate diagram (dashed lines) on the left-hand side does not exhibit a collision event. As shown below, this implies that the energies of both particles remain unchanged. 

We will now demonstrate how to perform the trace over the undetected particle for inelastic and elastic collisions, respectively. The result is represented on the right-hand side of Fig.~\ref{fig:trace1}.

\subsubsection*{Inelastic collisions.}

The complete expression for the inelastic scattering diagram, Fig.~\ref{fig:trace1}a), is given in Eq.~(\ref{eq:figtrace1a}). Focusing on those terms which are relevant for the trace over the undetected particle, this trace can be written in the following general form:
\begin{eqnarray}
& & \int_{-\infty}^\infty \frac{{\rm d}E{\rm d}E'}{|2\pi i|^2}\int\frac{{\rm d}{\bf k}}{(2\pi)^3}\bigl(\dots\bigr)^{(l)}_{(-E')}
\hat{G}_V^\dagger(E')|{\bf k}\rangle\langle{\bf k}|\hat{G}_V(E)\bigl(\dots\bigr)^{(r)}_{(-E)}\nonumber\\
& & = \int_{-\infty}^\infty \frac{{\rm d}E}{2\pi i}\bigl(\dots\bigr)^{(l)}_{(-E)} \left(\hat{G}_V^\dagger(E)-\hat{G}_V(E)\right)\bigl(\dots\bigr)^{(r)}_{(-E)}\,.
\label{eq:trace3}
\end{eqnarray}
On the left-hand side of Eq.~(\ref{eq:trace3}), ${\bf k}$ corresponds to  the final state of the undetected particle, whereas $\hat{G}_V(E)$ and $\hat{G}_V^\dagger(E')$ refer to the (single-particle) Green's operators expressing propagation from the collision event to the final state. According to the rules given in Sec.~\ref{sec:nparticle}, the collision events are associated with integrals $\int {\rm d}E/(-2\pi i)$ and $\int {\rm d}E'/(2\pi i)$ which determine the energies of the undetected ($E$ and $E'$) and the detected particle ($2E_i-E$ and $2E_i-E'$). The brackets $\bigl(\dots\bigr)^{(l)}_{(-E')}$ and $\bigl(\dots\bigr)^{(r)}_{(-E)}$ denote all the remaining parts of the scattering diagram where the energy argument enters with a {\em negative} sign, see Eq.~(\ref{eq:figtrace1a}). Their precise form is irrelevant for Eq.~(\ref{eq:trace3}) -- except for the fact that
$\bigl(\dots\bigr)^{(l)}_{(-E')}$ is a complex analytic function with poles only in the lower half of the complex plane, and 
$\bigl(\dots\bigr)^{(r)}_{(-E)}$ in the upper half. Due to the negative sign, this is in contrast to the respective contributions $\hat{G}_V^\dagger(E')$ and $\hat{T}_U^\dagger (E')$, as well as $\hat{G}_V(E)$ and $\hat{T}_U(E)$, which exhibit poles only in the {\em upper} (or lower) half plane.

\begin{figure}
\centerline{\includegraphics[width=12cm]{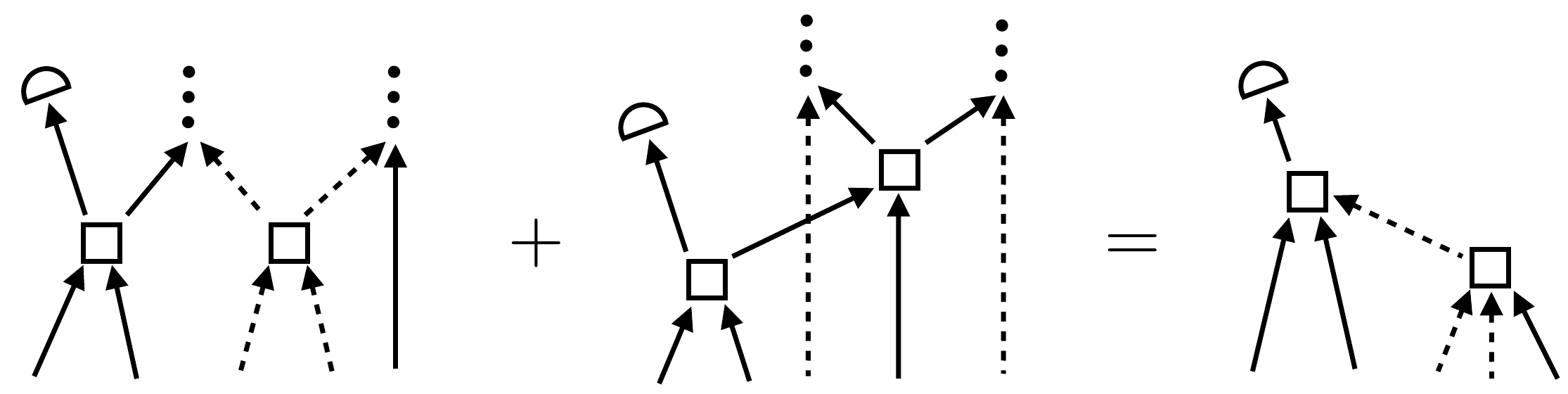}}
\caption{Elastic scattering diagram where the conjugate undetected amplitude originates from a previous elastic scattering event.
The sum of the two processes shown on the left-hand side reproduces the Gross-Pitaevskii diagram (cf. \cite{wellens09c}) on the right-hand side.}
\label{fig:trace2}
\end{figure}

Under these conditions -- which do not only hold for the example shown in Fig.~\ref{fig:trace1}, but for all other inelastic scattering diagrams we will encounter in the following -- the result of the trace is given on the right-hand side of Eq.~(\ref{eq:trace3}).
This general formula is proven in \ref{sec:trace}. Graphically, the result is depicted on the right-hand side of Fig.~\ref{fig:trace1}a). As a consequence, the energies
$E$ and $E'$ are set equal to each other, and the two conjugate Green's functions $\hat{G}_V^\dagger(E)$ and $\hat{G}_V(E)$ are replaced
by their difference $\left[\hat{G}_V^\dagger(E)-\hat{G}_V(E)\right]/(2\pi i)$ (which is also known as the \lq spectral function\rq, since the imaginary part of the Green's function determines the density of states \cite{spectral}).

\subsubsection*{Elastic collisions.}

In a similar way, the trace in the elastic scattering diagram, see Eq.~(\ref{eq:figtrace1b}), is performed as follows:
\begin{eqnarray}
& & \int_{-\infty}^\infty \frac{{\rm d}E}{2\pi i}\int\frac{{\rm d}{\bf k}}{(2\pi)^3}
\langle{\bf k}_i|\left(\hat{\Omega}_+^{(V)}(E_i)\right)^\dagger|{\bf k}\rangle\langle{\bf k}| \hat{G}_V(E) \bigl(\dots\bigr)_{E}\nonumber\\
& & =\langle{\bf k}_{i}|\left(\hat{\Omega}_+^{(V)}(E_i)\right)^\dagger\bigl(\dots\bigr)_{E_i}
\label{eq:trace4}\,,
\end{eqnarray}
see \ref{sec:trace}. According to Eq.~(\ref{eq:trace4}) -- which is graphically depicted in Fig.~\ref{fig:trace1}b) -- the outgoing solid arrow emitted from the two-body collision event is replaced by an incoming dashed arrow with energy $E_i$. We note that  precisely this diagram is the only interaction contribution generated by the Gross-Pitaevskii equation \cite{wellens09c}. Thereby, we have shown that our $N$-particle scattering theory reproduces the Gross-Piatevskii equation if only elastic scattering is taken into account.

In Eq.~(\ref{eq:trace4}), the conjugate undetected particle originates directly from the initial state $\langle {\bf k}_i|$ propagated in the disorder potential through the M\o ller operator $\left(\hat{\Omega}_+^{(V)}(E_i)\right)^\dagger$. The formula can be generalized, however, to the case where the
undetected particle undergoes previous collisions with other particles before colliding with the detected particle. An example is depicted in Fig.~\ref{fig:trace2}. Also in this case, the corresponding Gross-Pitaevskii diagram is reproduced (i.e. the outgoing solid arrow is replaced by an incoming dashed arrow). In a similar way, also the inelastic trace formula, Eq.~(\ref{eq:trace3}), is valid in the case where the undetected particle undergoes further collisions with other particles before the trace is taken -- provided that none of these other particles, in turn,  collides with the detected particle which, as discussed in Sec.~\ref{sec:incoherent}, is the case for a weak disorder potential. This allows us to take the trace over the undetected particles directly after their last collision with the detected particle -- without being obliged to follow their further evolution before finally leaving the scattering region. 

\section{Incoherent transport}
\label{sec:incoherent}

\subsection{Ladder diagrams}
\label{sec:ladder}

 \begin{figure}
\centerline{\includegraphics[width=10cm]{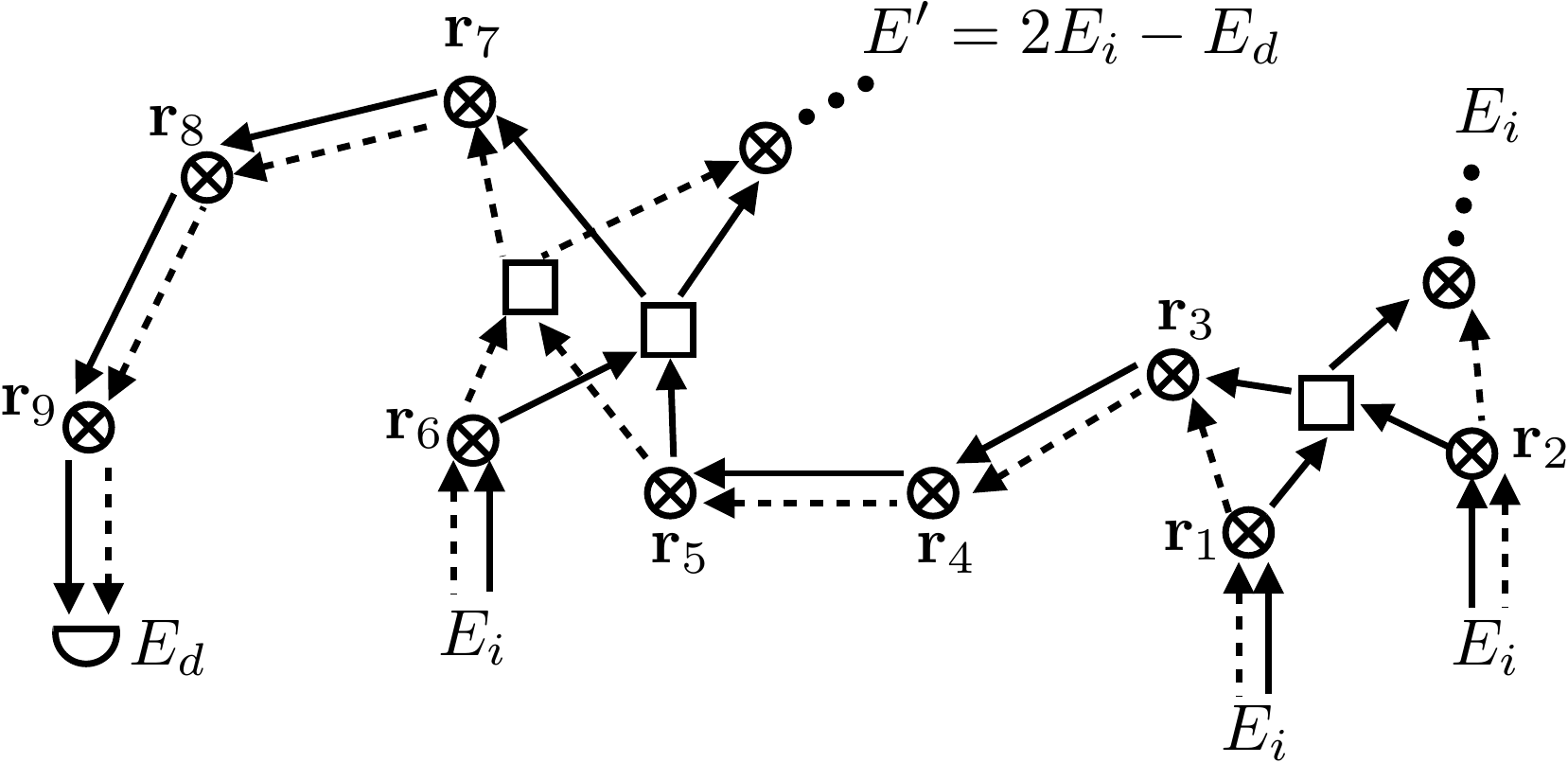}}
\caption{Example of a ladder diagram describing the propagation of three interacting particles 
in a slab with a random scattering potential. Pairs of conjugate amplitudes (solid and dashed arrows, respectively) undergo the same sequence of scattering events (encircled crosses) 
induced by the disorder potential, see Eq.~(\ref{eq:vcorr}), at ${\bf r}_1,\dots, {\bf r}_9$. Due to particle-particle collision events (squares), the particles redistribute their energies. Here, 
solid and dashed arrows correspond to disorder averaged single-particle Green's functions, Eq.~(\ref{eq:gav}), and their complex conjugates, respectively. Upon flux detection,
one particle is annihilated, 
while the undetected particles are traced over (dots).}
\label{fig:ladder}
\end{figure}

The $N$-particle scattering formalism outlined above is valid for an arbitrary potential $V({\bf r})$. Now, we 
consider $V({\bf r})$ as a random potential, and calculate the corresponding average density matrix $\overline{|f_+\rangle\langle f_+|}$.
For this purpose, we assume a Gaussian white noise potential, specified by the mean value $\langle V({\bf r})\rangle=0$ and
the two-point correlation function:
\begin{equation}
\overline{V({\bf r}_1)V({\bf r}_2)}=\frac{4\pi}{\ell_{\rm dis}}\delta({\bf r}_1-{\bf r}_2)\label{eq:vcorr}\,.
\end{equation}
Furthermore, the disorder potential is assumed to be weak, i.e. $\sqrt{E}\ell_{\rm dis}\gg 1$ for all relevant single-particle energies 
$E$, see Eq.~(\ref{eq:energy}). Initially, this is the case if $\sqrt{E_i}\ell_{\rm dis}\gg 1$. Due to inelastic collisions, the energies will change, but, as we will see later, their distribution will still be centered close to $E_i$, with only a negligible fraction of particles that reach single-particle energies $E\simeq 0$.

For the case of a single particle, the
disorder average in the limit $k\ell_{\rm dis}\gg 1$  is well known \cite{rammer,rossum99}: first, the vacuum Green's function $\hat{G}_0$, see Eq.~(\ref{eq:g0}), is replaced by the
average single-particle Green's function:
\begin{equation}
\langle {\bf k}'|\overline{\hat{G}}(E)|{\bf k}\rangle=(2\pi)^3 \delta({\bf k}-{\bf k}') G_E(k)\label{eq:gav}\,,
\end{equation}
with
\begin{equation}
 G_E(k)=\frac{1}{\tilde{k}_E^2-k^2}\label{eq:gav2}\,,
 \end{equation}
 where $\tilde{k}_E=\sqrt{E}+i/(2\ell_{\rm dis})$. In position representation, 
 this leads to an exponential decay of the average density with $2~{\rm Im}\tilde{k}_E=1/\ell_{\rm dis}$ as the decay constant, see Eq.~(\ref{eq:avint}) below. 
This establishes $\ell_{\rm dis}$ as the mean free path, i.e. the average distance between subsequent disorder scattering events.
 Second, when calculating the average density matrix $\overline{|f_+\rangle\langle f_+|}$,
 and representing both $|f_+\rangle$ and $\langle f_+|$ as a sum of diagrams, only those combination of diagrams survive where both $|f_+\rangle$ and $\langle f_+|$ undergo the same sequence of disorder scattering events. Here, a disorder scattering event is induced by the
 correlation function, Eq.~(\ref{eq:vcorr}), where $V({\bf r}_1)$ acts in $|f_+\rangle$ and
 $V({\bf r}_2)$ in  $\langle f_+|$ (or vice versa -- whereas correlators with $V({\bf r}_1)$ and $V({\bf r}_2)$ both acting in  $|f_+\rangle$ or both in $\langle f_+|$ are accounted for by the average Green's function (\ref{eq:gav2}) \cite{rammer}). These combinations of diagrams give rise to so-called ladder diagrams for the average density \cite{rossum99}.
 
\begin{figure}
\centerline{\includegraphics[width=6.5cm]{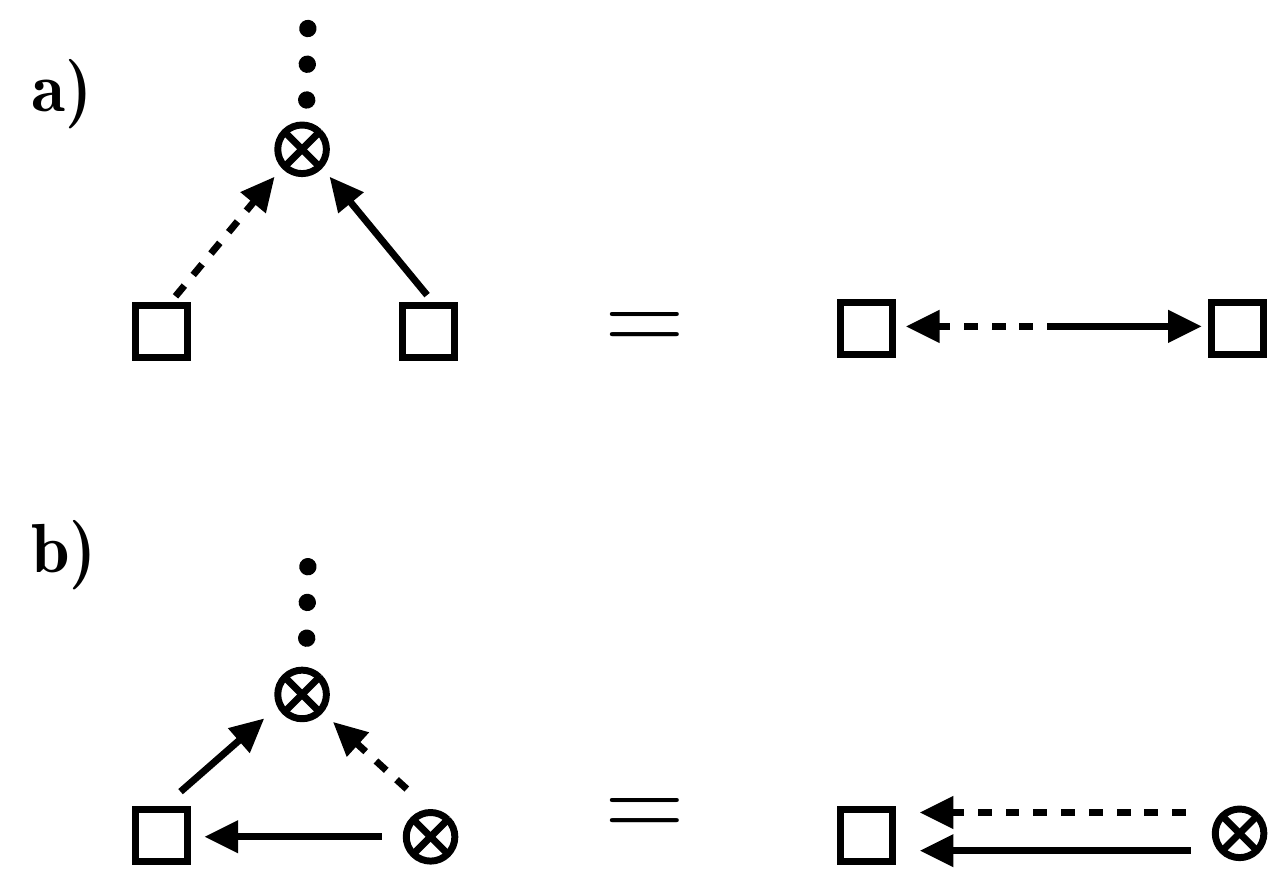}}
\caption{Trace over the undetected particle for disorder-averaged diagrams (see Fig.~\ref{fig:ladder}), in case of {\bf a)} inelastic or {\bf b)} elastic collisions.
In contrast to Fig.~\ref{fig:trace1}, arrows refer to the disorder-averaged Green's function, Eq.~(\ref{eq:gav2}). The solid-dashed double arrow in {\bf a)} denotes the average spectral function $[G_E^*(k)-G_E(k)]/(2\pi i)$.
}
\label{fig:trace3}
\end{figure}

 We now apply the same procedure to the $N$-particle scattering processes presented in Sec.~\ref{sec:nparticle}. 
 First, we take a diagram contributing to $|f_+\rangle$ and another one (called \lq conjugate diagram\rq\ in the following) contributing to 
 $\langle f_+|$. Then, we replace all vacuum Green's functions by average Green's functions and correlate, using Eq.~(\ref{eq:vcorr}), each disorder scattering event  with another one in the conjugate diagram such that  both conjugate diagrams undergo the same sequence of disorder scattering events. Finally, we choose one of the final particles 
 as detected particle, and trace over the remaining $N-1$ particles, see Eq.~(\ref{eq:trace}). An example is shown in Fig.~\ref{fig:ladder}. 
 In this figure, the trace over undetected particles is performed as soon as the corresponding particle (solid line)  is re-united with its conjugate counterpart (dashed line) at a disorder scattering event. It turns out that the same result is obtained if the trace is performed  before taking the disorder average  according to Fig.~\ref{fig:trace1}. This leads to disorder-averaged trace formulas as depicted in Fig.~\ref{fig:trace3}, which we will use in the following to evaluate the trace over the undetected particles.
 
Among the $N$-particle ladder diagrams thus constructed, we neglect all those where two particles which interacted once meet again. This approximation is equivalent to the neglect of recurrent scattering \cite{recurrent} for a single particle,
 which, alike the neglect of non-ladder diagrams, is valid for $k\ell_{\rm dis}\gg 1$. It allows 
 us to trace away the undetected particles directly after their interaction with the detected particle.
Finally, we assume that at least one disorder scattering event occurs between two collision events. This is justified if $\ell_{\rm int}\gg\ell_{\rm dis}$ where
\begin{equation}
\ell_{\rm int}=\frac{1}{\sigma\rho_0}\,,
\end{equation}
with $\sigma$ denoting the scattering cross section of the atom-atom interaction potential $U({\bf r})$,
defines the average distance between two inelastic collision events. For $s$-wave scattering, $\sigma=8\pi a_s^2$, see Eq.~(\ref{eq:lintswave}).

\subsection{Building blocks}

\begin{figure}
\centerline{\includegraphics[width=13cm]{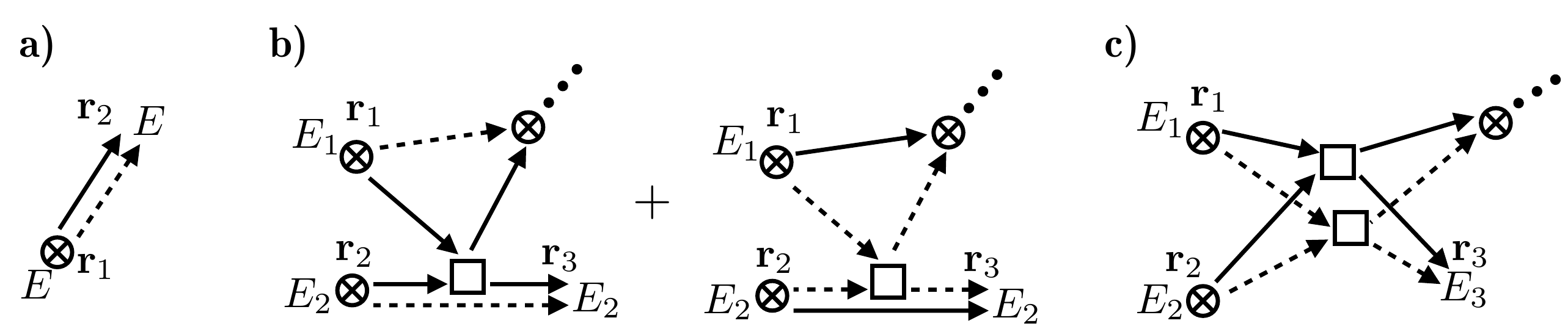}}
\caption{The three building blocks from which all ladder diagrams (see Fig.~\ref{fig:ladder}) are constructed. {\bf a)} Single-particle propagation in the disorder potential, see Eq.~(\ref{eq:avint}). {\bf b)} Elastic two-particle collision $g_{E_1,E_2}({\bf r}_1,{\bf r}_2,{\bf r}_3)$, see Eq.~(\ref{eq:g}). 
{\bf c)} Inelastic two-particle collision $f_{E_1,E_2,E_3}({\bf r}_1,{\bf r}_2,{\bf r}_3)$, see Eq.~(\ref{eq:f}).}
\label{fig:ladderbb}
\end{figure}

The trace over the undetected particle allows us to decompose every ladder diagram (like the one shown in Fig.~\ref{fig:ladder}) into independent building blocks. These building blocks are shown in Fig.~\ref{fig:ladderbb}.
The first one, Fig.~\ref{fig:ladderbb}a), represents a single average propagation step of a single particle with energy $E$ and corresponding wave vector ${\bf k}$ in the disordered potential from ${\bf r}_1$ to ${\bf r}_2$:
\begin{equation}
P_E({\bf r}_1,{\bf r}_2)=\frac{4\pi}{\ell_{\rm dis}}\left|\int\frac{{\rm d}{\bf k}}{(2\pi)^3}~e^{-i {\bf k}\cdot({\bf r}_{1}-{\bf r}_2)} G_E(k)\right|^2=\frac{e^{-|{\bf r}_1-{\bf r}_2|/\ell_{\rm dis}}}{4\pi\ell_{\rm dis}|{\bf r}_1-{\bf r}_2|^2}\label{eq:avint}\,,
\end{equation}
for $E\geq 0$.  Note that, for a white noise potential as defined in Eq.~(\ref{eq:vcorr}), the mean free path $\ell_{\rm dis}$ is independent of $E$ \cite{akkermans}. 
For $E<0$, the propagation is exponentially suppressed; in this case, Eq.~(\ref{eq:avint}) is multiplied by an additional factor $\exp\left(-2
|{\bf r}_1-{\bf r}_2|\sqrt{|E|}\right))$. Since the typical distance between two scattering events is given by the mean free path $\ell_{\rm dis}$, we can neglect the occurrence of negative energies if $\sqrt{|E|}\ell_{\rm dis}\gg 1$.

The second building block, Fig.~\ref{fig:ladderbb}b), represents an elastic collision event, where the energies of both particles 
are unchanged:
\begin{eqnarray}
& & g_{E_1,E_2}({\bf r}_1,{\bf r}_2,{\bf r}_3)=2~\left(\frac{4\pi}{\ell_{dis}}\right)^2~2 {\rm Re}\Biggl\{\frac{1}{2}
\int\frac{{\rm d}{\bf k}_1\dots {\rm d}{\bf k}_5}{(2\pi)^{15}}
\nonumber\\
& & \times e^{-i[({\bf k}_{1}-{\bf k}_4)\cdot{\bf r}_{1}+({\bf k}_2-{\bf k}_5)\cdot{\bf r}_2+({\bf k}_{5}-{\bf k}_{3})\cdot{\bf r}_{3}]}\langle{\bf k}_3,{\bf k}_4|\hat{T}_U(E_1+E_2)|{\bf k}_1,{\bf k}_2\rangle\nonumber\\
& & \times G_{E_1}(k_1)G_{E_2}(k_2)G_{E_2}(k_3)G^*_{E_1}(k_4)G^*_{E_2}(k_5)
\Biggr\}\,. \label{eq:g}
\end{eqnarray}
The trace over the undetected particle was performed according to Fig.~\ref{fig:trace3}b), giving rise to an average Green's function $G^*_{E_1}(k_4)$. For reasons of clarity, the wave vectors 
${\bf k}_1,\dots,{\bf k}_5$ are not explicitly shown in Fig.~\ref{fig:ladderbb}. They can, however, be easily deduced from the phase factors
$\exp(\pm i {\bf k}\cdot {\bf r})$ describing annihilation or creation of a particle ${\bf k}$ due to disorder scattering at ${\bf r}$, see Eqs.~(\ref{eq:vsingle},\ref{eq:v}),  with the help of following rule: outgoing solid (dashed) arrows always contribute with negative (positive) sign, the opposite holds for incoming arrows.
For example, ${\bf k}_5$  -- with phase factor $\exp[i{\bf k}_5\cdot({\bf r}_2-{\bf r}_3)]$ in Eq.~(\ref{eq:g}) -- is associated with the dashed arrow pointing from ${\bf r}_2$ to ${\bf r}_3$ in Fig.~\ref{fig:ladderbb}b).

The first factor 2 in Eq.~(\ref{eq:g}) originates from the fact that  the solid and dashed  incoming amplitudes can be grouped  together in two different ways. It can be shown that this accounts for fluctuations of the atomic density inside the disordered slab \cite{wellens09b}. 
The factor $1/2$ in front of the integral originates from the indistinguishability of particles, see the discussion at the end of \ref{sec:examplediag}.

Finally, the third building block, Fig.~\ref{fig:ladderbb}c), amounts to an inelastic  collision event, where the energies of two particles $E_1$ and $E_2$ change to $E_3$ and $E_4=E_1+E_2-E_3$:
\begin{eqnarray}
& & f_{E_1,E_2,E_3}({\bf r}_1, {\bf r}_2,{\bf r}_3)  =  2~\left(\frac{4\pi}{\ell_{\rm dis}}\right)^2
\int\frac{{\rm d}{\bf k}_4}{(2\pi)^3} \frac{G^*_{E_1+E_2-E_3}(k_4)-G_{E_1+E_2-E_3}(k_4)}{2\pi i}
\nonumber\\
& & \times\Biggl| \frac{1}{2}\int\frac{{\rm d}{\bf k}_1{\rm d}{\bf k}_2{\rm d}{\bf k}_3}{(2\pi)^{9}} 
e^{-i({\bf k}_{1}\cdot{\bf r}_1+{\bf k}_{2}\cdot{\bf r}_{2}-{\bf k}_{3}\cdot {\bf r}_{3})} 
\langle{\bf k}_3,{\bf k}_4|\hat{T}_U(E_1+E_2)|{\bf k}_1,{\bf k}_2\rangle
\nonumber\\
& & \times G_{E_1}(k_1)G_{E_2}(k_2)G_{E_3}(k_3)\Biggr|^2 \,, \label{eq:f}
\end{eqnarray}
where Fig.~\ref{fig:trace3}a) was used for the trace over the undetected particle. 

\subsection{Transport equation}

The outgoing arrows of each building block may now be attached to the incoming arrows of the next building block, and so on. The sum of ladder diagrams
resulting from all combinations of these building blocks is expressed by the following nonlinear integral equation:
\begin{eqnarray}
I_E({\bf r}) & = & I_0({\bf r})\delta(E-E_i)+\int_{\mathcal V}{\rm d}{\bf r}'P_E({\bf r},{\bf r}')I_E({\bf r}')\nonumber\\
& & +\int_{0}^\infty {\rm d}E'\int_{\mathcal V}{\rm d}{\bf r}'\int_{\mathcal V}{\rm d}{\bf r}'' \biggl[g_{E',E}({\bf r}',{\bf r}'',{\bf r}) I_E({\bf r}'')\biggr.\nonumber\\
& & +\biggr.\int_{0}^\infty {\rm d}E''  f_{E',E'',E}({\bf r}',{\bf r}'',{\bf r}) I_{E''}({\bf r}'')\biggl]I_{E'}({\bf r}')\label{eq:transport}\,,
\end{eqnarray}
where
\begin{equation}
I_0({\bf r})=\rho_0e^{-z_{{\bf r},-\hat{\bf k}_i}/\ell_{\rm dis}}\label{eq:i0}
\end{equation}
represents the incoming wave propagating to ${\bf r}$ without being scattered. Correspondingly, $z_{{\bf r},-\hat{\bf k}_i}$ denotes the distance 
from the surface of the scattering region ${\mathcal V}$ to ${\bf r}$ along a straight line parallel to the direction ${\bf k}_i$ of the incident wavepacket.
The quantity 
$I_E({\bf r})$ can be interpreted as the average density of particles with energy $E$ at position ${\bf r}$ (at least in the case $\ell_{\rm dis}\sqrt{E}\gg 1$ of weak disorder where
the spectral function $[G^*_E(k)-G_E(k)]/(2\pi i)\to \delta(E-k^2)$ approaches a $\delta$-function, such that a particle with wavevector ${\bf k}$ possesses a well defined energy). In particular,
$I({\bf r})=\int{\rm d}E I_E({\bf r})=\overline{\langle f_+|\hat{\rho}({\bf r})|f_+\rangle}$ gives the disorder-averaged expectation value of the single-particle density operator
$\hat{\rho}({\bf r})=\hat{\psi}^\dagger({\bf r})\hat{\psi}({\bf r})$ with respect to the quasi-stationary scattering state $|f_+\rangle$.
From $I_E({\bf r})$, the diffuse flux $\gamma^{(L)}(\hat{\bf k}_d)$ --  i.e. the disorder average of $\gamma(\hat{\bf k}_d)$, see Eq.~(\ref{eq:gamma}), 
in ladder approximation -- of particles scattered into direction $\hat{\bf k}_d$ is finally obtained as:
\begin{equation}
\gamma^{(L)}(\hat{\bf k}_d)=\int_0^\infty{\rm d}E~ \gamma_E^{(L)}(\hat{\bf k}_d)\label{eq:jsc}\,,
\end{equation}
where 
\begin{equation}
\gamma_E^{(L)}(\hat{\bf k}_d)=\int_{\mathcal V}\frac{{\rm d}{\bf r}}{{\mathcal A}\ell_{\rm dis}}~
e^{-z_{{\bf r},\hat{\bf k}_d}/\ell_{\rm dis}}
\sqrt{\frac{E}{E_i}} \frac{I_E({\bf r})}{\rho_0}\label{eq:jsce}
\end{equation}
denotes the ladder component of the average spectral flux density, i.e. the flux of particles scattered into direction $\hat{\bf k}_d$ with energy $E$, see Eq.~(\ref{eq:gammae}). In Eq.~(\ref{eq:jsce}), $z_{{\bf r},\hat{\bf k}_d}$ denotes the distance from ${\bf r}$ to the surface of the scattering region in direction $\hat{\bf k}_d$. Note that, in the far-field limit, only positive energies contribute to the scattered flux, Eq.~(\ref{eq:jsc}). Within the scattering medium, negative energies are neglected in the transport equation (\ref{eq:transport}) due to the exponential suppression mentioned after Eq.~(\ref{eq:avint}).

Since we assume that disorder scattering events -- represented by the term $P_E$ in Eq.~(\ref{eq:transport}) --  are much more frequent than collision events, we may neglect the spatial dependence of the collision terms in Eq.~(\ref{eq:transport}) and approximate them by $\delta$-functions: $g_{E';E}({\bf r}',{\bf r}'',{\bf r}) \simeq \delta({\bf r}'-{\bf r})\delta({\bf r}''-{\bf r})g_{E';E}$
and $f_{E',E'',E}({\bf r}',{\bf r}'',{\bf r}) \simeq \delta({\bf r}'-{\bf r})\delta({\bf r}''-{\bf r})f_{E',E'',E}$,
where
\begin{eqnarray}
g_{E',E} & = & \int{\rm d}{\bf r}'{\rm d}{\bf r}''~g_{E',E}({\bf r}',{\bf r}'',{\bf r})\label{eq:gcontact}\,,\\
f_{E',E'',E} & = & \int{\rm d}{\bf r}'{\rm d}{\bf r}''~f_{E',E'',E}({\bf r}',{\bf r}'',{\bf r})\label{eq:fcontact}\,.
\end{eqnarray}
The transport equation (\ref{eq:transport}) then reduces to:
\begin{eqnarray}
I_E({\bf r}) & = & I_0({\bf r})\delta(E-E_i)+\int_{\mathcal V}{\rm d}{\bf r}'P_E({\bf r},{\bf r}')I_E({\bf r}')\nonumber\\
& & +\int_{0}^\infty {\rm d}E' \left[g_{E',E} I_E({\bf r})+\int_{0}^\infty {\rm d}E''  f_{E',E'',E} I_{E''}({\bf r})\right]I_{E'}({\bf r})\,.
\label{eq:transport2}
\end{eqnarray}
As shown in \cite{schwiete13b}, this equation can also be derived from the nonlinear Boltzmann transport equation.
Due to the above collision approximation, the spatial transport of particles in Eq.~(\ref{eq:transport2}) is solely governed by the propagation $P$ in the disorder potential, whereas the collision terms $g$ and $f$ lead to a redistribution of energies. As compared to Eqs.~(\ref{eq:g},\ref{eq:f}), these terms simplify as follows:
\begin{eqnarray}
g_{E_1,E_2}& = & \left(\frac{4\pi}{\ell_{\rm dis}}\right)^2 
\int\frac{{\rm d}{\bf k}_1{\rm d}{\bf k}_2}{(2\pi)^{6}} \left|G_{E_1}(k_1)\right|^2~2\,{\rm Re}\left\{\langle{\bf k}_{12}|\hat{T}^{(1)}_U(E_{12})|{\bf k}_{12}\rangle \right.\nonumber\\
& & \times\left.\left|G_{E_2}(k_2)\right|^2 G_{E_2}(k_2) \right\}\label{eq:gsimple}\,,\\
f_{E_1,E_2,E_3} & = &  2~\left(\frac{4\pi}{\ell_{\rm dis}}\right)^2~\frac{1}{4} 
\int\frac{{\rm d}{\bf k}_1{\rm d}{\bf k}_2{\rm d}{\bf k}_3}{(2\pi)^{9}} 
\frac{G^*_{E_1+E_2-E_3}(k_4)-G_{E_1+E_2-E_3}(k_4)}{2\pi i}
\nonumber\\
& & \times \left|\langle{\bf k}_{34}|\hat{T}^{(1)}_U(E_{12})|{\bf k}_{12}\rangle\right|^2
\left|G_{E_1}(k_1)\right|^2 \left|G_{E_2}(k_2)\right|^2 \left|G_{E_3}(k_3)\right|^2\label{eq:fsimple}\,,
\end{eqnarray}
with ${\bf k}_4={\bf k}_1+{\bf k}_2-{\bf k}_3$, $E_{12}=E_1+E_2-E_{{\bf k}_1+{\bf k}_2}/2$, and $|{\bf k}_{12}\rangle$, $|{\bf k}_{34}\rangle$ as defined after Eq.~(\ref{eq:tmatrix}).

\subsection{Thermalization}
\label{sec:thermalization}

For a given form of the two-body $T$-matrix (e.g. s-wave scattering, see below), we can now calculate the collision terms $g$ and $f$ according to Eqs.~(\ref{eq:gsimple},\ref{eq:fsimple}), and then
numerically solve the transport equation (\ref{eq:transport2}) by iteration. Before presenting the corresponding  numerical results in
Sec.~\ref{sec:results}, however, we will discuss, in the remainder of this section, some general properties of  $g$ and $f$, which, as shown below, lead to thermalization of the single-particle energies for an infinite system.

As shown in \ref{sec:cons}, the collision terms fulfill the following relations:
\begin{eqnarray}
\sqrt{E_2}g_{E_1;E_2} & = & -\int_0^{\infty} {\rm d}E~\sqrt{E} f_{E_1,E_2,E}\label{eq:particlecons}\,,\\
(E_1+E_2)\sqrt{E_2}g_{E_1;E_2}& = & -\int_0^\infty {\rm d}E~2E\sqrt{E} f_{E_1,E_2,E}\label{eq:energycons}\,.
\end{eqnarray}
Both relations follow from the fact that the $T$-matrix associated to the atom-atom interaction potential $U({\bf r})$ fulfills the optical theorem, Eq.~(\ref{eq:opttheorem}), and express conservation of the particle and the energy flux, respectively.
Moreover, from Eq.~(\ref{eq:f}), one can show that:
\begin{equation}
\frac{f_{E_1,E_2,E_3}}{\sqrt{E_1+E_2-E_3}} = \frac{f_{E_3,E_1+E_2-E_3,E_1}}{\sqrt{E_2}}\label{eq:reversibility}\,.
\end{equation}
This equation expresses microscopic reversibility of the collision dynamics: given two particles with energy $E_1$ and $E_2$, 
the collision process $E_1,E_2\to E_3,E_4$ occurs with the same probability as
the reverse process $E_3,E_4\to E_1,E_2$ given two particles with energy $E_3$ and $E_4$. The square roots in the denomimators of
(\ref{eq:reversibility}) result from the traces over the undetected particle with energy $E_4=E_1+E_2-E_3$ (left-hand side) 
or $E_2$ (right-hand side), respectively.

Using Eqs.~(\ref{eq:particlecons},\ref{eq:energycons}) -- and the fact that the linear propagator $P_E({\bf r},{\bf r}')=P({\bf r},{\bf r}')$ is independent of
$E$ -- it follows that the quantities $J({\bf r})=\int_0^\infty {\rm d}E~J_E({\bf r})$ and $K({\bf r})=\int_0^\infty {\rm d}E~K_E({\bf r})$, with
\begin{equation}
J_E({\bf r})=\sqrt{E}I_E({\bf r}),\ K_E({\bf r})=E\sqrt{E}I_E({\bf r})\label{eq:jk}\,,
\end{equation}
corresponding to the particle and energy flux, respectively,
both fulfill the same linear transport equation:
\begin{eqnarray}
J({\bf r}) & = &  J_0({\bf r})+\int_{\mathcal V}{\rm d}{\bf r}'P({\bf r},{\bf r}')J({\bf r}')\label{eq:jtot}\,,\\
K({\bf r})  & =  & K_0({\bf r})+\int_{\mathcal V}{\rm d}{\bf r}'P({\bf r},{\bf r}')K({\bf r}')\label{eq:ktot}\,,
\end{eqnarray}
where the source terms $J_0({\bf r})=\sqrt{E_i}I_0({\bf r})$ and $K_0({\bf r})=E_i\sqrt{E_i}I_0({\bf r})$ differ only by the constant factor $E_i$. 
Due to Eqs.~(\ref{eq:particlecons},\ref{eq:energycons}),
the collision terms drop out from Eq.~(\ref{eq:transport2}) when integrating over $E$. Since the linear transport equation fulfills flux conservation, this, in turn, implies that, both, particle and energy flux are conserved. Furthermore, since $K_0({\bf r})=E_i J_0({\bf r})$, the same relation holds for the solutions of the linear equations (\ref{eq:jtot},\ref{eq:ktot}):
\begin{equation}
K({\bf r})=E_i J({\bf r})\label{eq:KJ}\,.
\end{equation}
After these preparatory steps, we can now look for a solution of the transport equation (\ref{eq:transport2}) in case of a semi-infinite medium. Far away from its boundary, $I_E({\bf r})=I_E$ should become independent of $\bf r$, and $I_0({\bf r})$, Eq.~(\ref{eq:i0}), tends to zero. Hence, the constant solution $I_E$ must fulfill:
\begin{equation}
\int_{0}^\infty {\rm d}E' \left[g_{E',E} I_E+\int_{0}^\infty {\rm d}E''  f_{E',E'',E} I_{E''}\right]I_{E'}=0\label{eq:transportinf}\,.
\end{equation}
Using Eqs.~(\ref{eq:particlecons},\ref{eq:reversibility}), one can show that
$I_E=\sqrt{E}e^{-\gamma E}$
fulfills Eq.~(\ref{eq:transportinf}) for $\gamma>0$. The constant $\gamma$, in turn, is determined by Eq.~(\ref{eq:KJ}) as $\gamma=2/E_i$. Hence
the normalized particle flux distribution is given by:
\begin{equation}
\frac{J_E({\bf r})}{J({\bf r})}=\frac{4E}{E_i^2}e^{-2E/E_i}\label{eq:mb}\,.
\end{equation}
This corresponds to a Maxwell-Boltzmann distribution the temperature of which is determined by the initial energy ($E_i=k_B T/2$).
Thereby, we have demonstrated thermalization in case of a semi-infinite medium. In Sec.~\ref{sec:results}, we will study the transport behaviour predicted by Eq.~(\ref{eq:transport2}) for a finite medium, and see how the thermal distribution is approached during propagation through a finite slab.

\section{Coherent transport}
\label{sec:coherent}

\subsection{Crossed diagrams}

\begin{figure}
\centerline{\includegraphics[width=10cm]{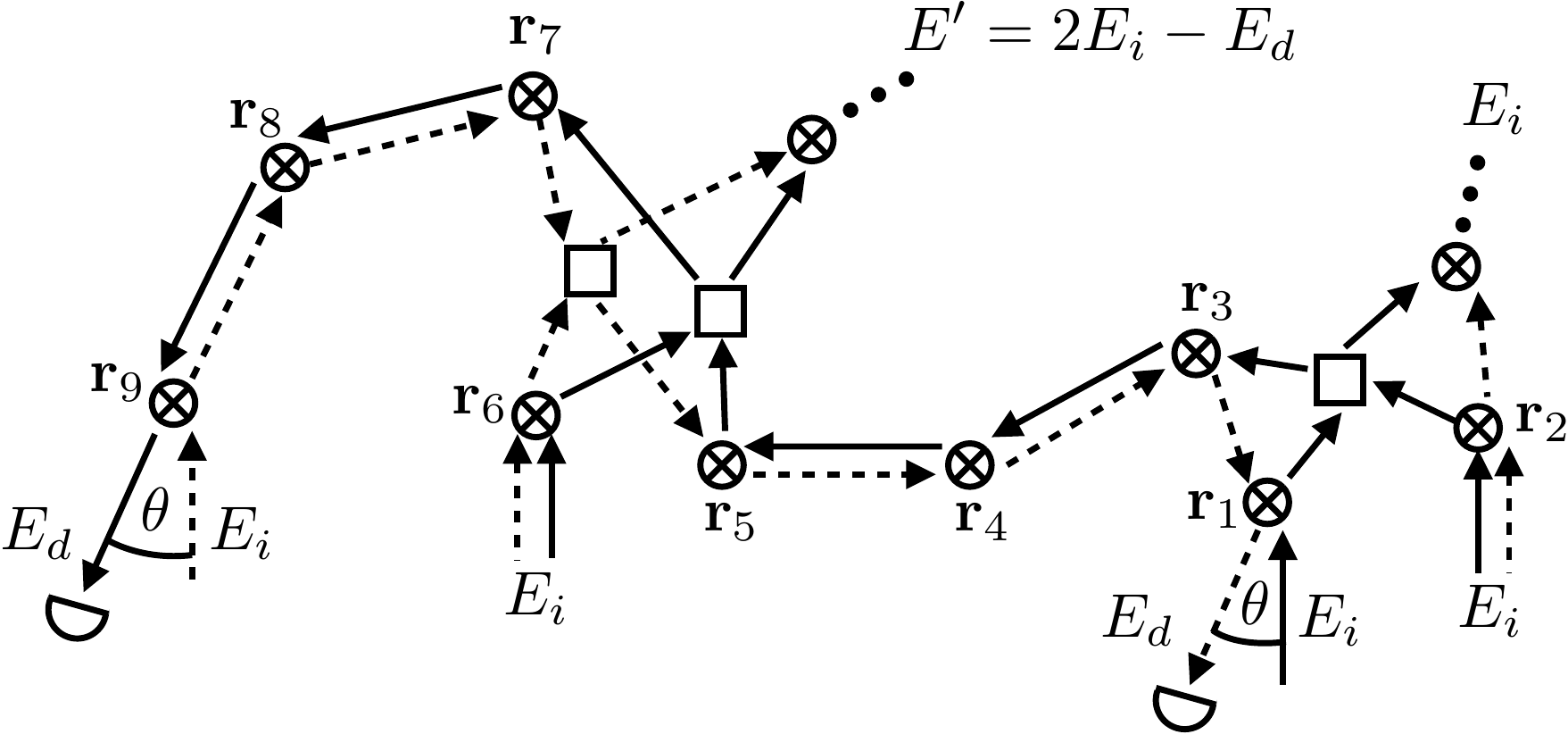}}
\caption{Example of a crossed diagram contributing to coherent backscattering  in the case of 
three interacting particles. It is obtained from the ladder diagram shown in Fig.~\ref{fig:ladder} by reversing the direction of propagation of the dashed propagators along the path ${\bf r}_1\to{\bf r}_3\to{\bf r}_4\to{\bf r}_5\to{\bf r}_7\to{\bf r}_8\to{\bf r}_9$.}
\label{fig:crossed}
\end{figure}

Before turning to the numerical results, however, 
we will extend the general formalism of Sec.~\ref{sec:incoherent} in order to calculate the leading interference correction (in the weak disorder parameter $1/(k\ell_{\rm dis})$) to the average scattered flux density. This correction is described by crossed diagrams \cite{langer66}, which are obtained from the ladder diagrams by reversing the direction of propagation of a single amplitude. Starting from the ladder diagram shown in Fig.~\ref{fig:ladder},
we can construct, for example, the crossed diagram shown in Fig.~\ref{fig:crossed}.
It amounts to an interference between two amplitudes where the detected atom is emitted from ${\bf r}_9$ and ${\bf r}_1$, respectively.
For a given wavevector ${\bf k}_d$ of the detected atom, the backscattering angle $\theta$ is defined by
 $\cos\theta=-\hat{\bf k}_i\cdot \hat{\bf k}_d$. Since annihilation and creation of atoms with wavevector ${\bf k}$ by the disorder potential at position ${\bf r}$ are associated with factors $e^{\pm i {\bf k}\cdot {\bf r}}$, respectively, see Eqs.~(\ref{eq:vsingle},\ref{eq:v}), this leads to a phase factor $e^{i{\bf q}\cdot ({\bf r}_1-{\bf r}_9)}$ with respect to the ladder diagram, Fig.~\ref{fig:ladder}, where
\begin{equation}
 {\bf q}={\bf k}_i+{\bf k}_d\label{eq:q}\,.
\end{equation}
Since ${\bf r}_1$ and ${\bf r}_9$ refer to randomly chosen positions of scattering events, this phase factor vanishes on average unless
${\bf q}\simeq 0\ \ \Leftrightarrow\ \ {\bf k}_i\simeq-{\bf k}_d$, corresponding to exact backscattering ($\theta=0$). Therefore, this effect of interference between reversed amplitudes  is called \lq coherent backscattering\rq\ \cite{kuga84,albada85,wolf85}. More precisely, one can show that the angular width of the coherent backscattering interference peak is approximately given by $\Delta\theta\simeq 1/(k\ell_{\rm dis})$ \cite{akkermans}.
For a single particle, the height of this peak at $\theta=0$ equals the incoherent background as described by the ladder diagrams (except for single scattering which only contributes to the background), what amounts to an enhancement of the backscattered flux by a factor 2. 
We will show below how this enhancement factor changes as a consequence of elastic and inelastic atom-atom collisions.

For this purpose, we will derive a transport equation for the \lq crossed density\rq\ which describes a pair of amplitudes propagating 
in opposite directions. In Fig.~\ref{fig:crossed}, the corresponding crossed scattering path is given by 
${\bf r}_1\to {\bf r}_3\to {\bf r}_4\to {\bf r}_5\to {\bf r}_7\to {\bf r}_8\to {\bf r}_9$ (where we define the direction of the path to be fixed by the solid arrows, whereas the dashed arrows propagate in the opposite sense). The remaining parts (${\bf r}_2$ and ${\bf r}_6$) correspond  to ladder diagrams already treated in Sec.~\ref{sec:incoherent}. 
Due to energy conservation, the energies $E$ and $\widetilde{E}$ associated with the two counterpropagating  conjugate amplitudes always fulfill the following relation:
\begin{equation}
\widetilde{E}=E_i+E_d-E\label{eq:etilde}\,.
\end{equation}

\subsection{Crossed building blocks}
\label{sec:crossedbb}

\begin{figure}
\centerline{\includegraphics[width=13cm]{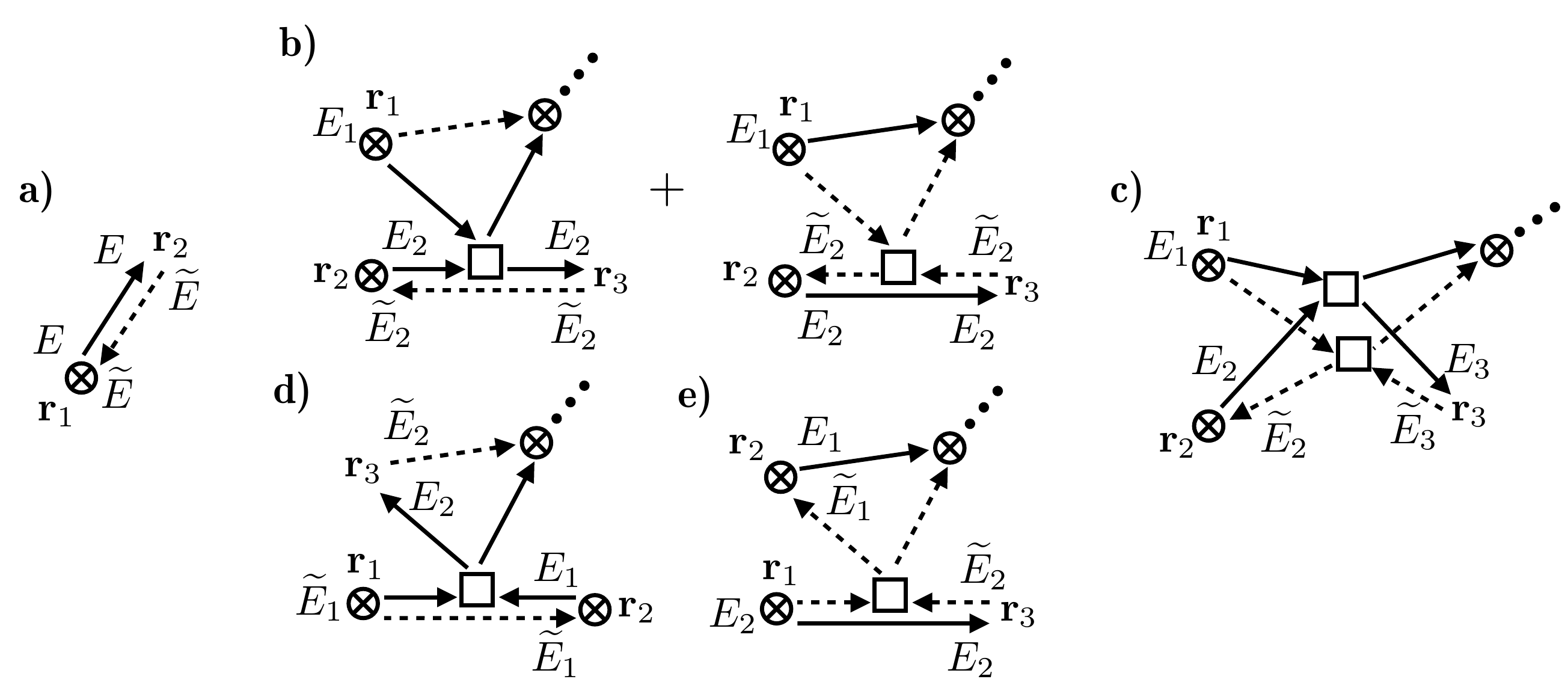}}
\caption{Building blocks for crossed diagrams.
{\bf a)} Single-particle propagation $P^{(C)}_E({\bf r}_1,{\bf r}_2)$, see Eq.~(\ref{eq:avintc}). Note that the energies $E$ and $\widetilde{E}$ associated to counterpropagating amplitudes (solid and dashed line, respectively) fulfill $E+\widetilde{E}=E_i+E_d$, see Eq.~(\ref{eq:etilde}). {\bf b)} Elastic collision $g^{(C)}_{E_1,E_2}$, see Eq.~(\ref{eq:gc}), obtained by reversing the lowermost line of the ladder building block $g_{E_1,E_2}$, Fig.~\ref{fig:ladderbb}b). {\bf c)} Inelastic collision $f^{(C)}_{E_1,E_2,E_3}$, see Eq.~(\ref{eq:fc}). {\bf d)} Crossed collision $h^{(C)}_{E_1,E_2}$, see Eq.~(\ref{eq:hc}),
obtained by reversing the solid arrow between ${\bf r}_1$ and ${\bf r}_3$ for $g_{E_1,E_2}$, Fig.~\ref{fig:ladderbb}b). {\bf e)} Conjugate crossed collision $\left(h^{(C)}\right)^*_{\widetilde{E}_2,\widetilde{E}_1}$. Note that, due to different possibilities of reversing single-particle amplitudes, there are more crossed than ladder building blocks (see Fig.~\ref{fig:ladderbb}).}
\label{fig:crossedbb}
\end{figure}

This leads us to the first crossed building block:
\begin{eqnarray}
P^{(C)}_E({\bf r}_1,{\bf r}_2)& = & \frac{4\pi}{\ell_{\rm dis}}\int\frac{{\rm d}{\bf k}_1{\rm d}{\bf k}_2}{(2\pi)^6}~e^{-i ({\bf k}_1-{\bf k}_2)\cdot({\bf r}_{1}-{\bf r}_2)} G_E(k_1)G^*_{\widetilde{E}}(k_2)\nonumber\\
& = & \frac{e^{|{\bf r}_1-{\bf r}_2|\left(i\sqrt{E}-i\sqrt{\widetilde{E}}-1/\ell_{\rm dis}\right)}}{4\pi\ell_{\rm dis}|{\bf r}_1-{\bf r}_2|^2}\,,
\label{eq:avintc}
\end{eqnarray}
describing single-particle propagation with different energies $E$ (wave vector ${\bf k}_1$) and $\widetilde{E}$ (wave vector ${\bf k}_2$) for the conjugate amplitudes, see Fig.~\ref{fig:crossedbb}a). For $E=\widetilde{E}$, i.e.
$E=(E_i+E_d)/2$ due to Eq.~(\ref{eq:etilde}), it reduces to the ladder propagator $P_E$, see Eq.~(\ref{eq:avint}). 

The following building block, Fig.~\ref{fig:crossedbb}b)\,,
\begin{eqnarray}
g^{(C)}_{E_1,E_2}& = & \left(\frac{4\pi}{\ell_{\rm dis}}\right)^2 
\int\frac{{\rm d}{\bf k}_1{\rm d}{\bf k}_2}{(2\pi)^{6}} \left|G_{E_1}(k_1)\right|^2 G_{E_2}(k_2) G_{\widetilde{E}_2}^*(k_2)
\label{eq:gc}\\
& & \times \Biggl[ G_{E_2}(k_2)\langle{\bf k}_{12}|\hat{T}^{(1)}_U(E_{12})|{\bf k}_{12}\rangle 
+G^*_{\widetilde{E}_2}(k_2)\langle{\bf k}_{12}|\left(\hat{T}^{(1)}_U(E'_{12})\right)^\dagger|{\bf k}_{12}\rangle \Biggr]\nonumber\,,
\end{eqnarray}
where $E_{12}=E_1+E_2-E_{{\bf k}_1+{\bf k}_2}/2$, $E'_{12}=E_1+\widetilde{E}_2-E_{{\bf k}_1+{\bf k}_2}/2$, and $|{\bf k}_{12}\rangle$ as defined after Eq.~(\ref{eq:tmatrix}), represents the crossed counterpart of the elastic collision $g_{E_1,E_2}$, see Fig.~\ref{fig:ladderbb}b). Again, it reduces to the corresponding ladder term, Eq.~(\ref{eq:gsimple}), for $\widetilde{E}_2=E_2$.
The wavevector ${\bf k}_1$ in Eq.~(\ref{eq:gc}) is associated with Green's functions emitted from position ${\bf r}_1$, and ${\bf k}_2$ with those propagating between ${\bf r}_2$ and ${\bf r}_3$. 

Similarly, Fig.~\ref{fig:crossedbb}c),
\begin{eqnarray}
f^{(C)}_{E_1,E_2,E_3} & = &  4~\left(\frac{4\pi}{\ell_{\rm dis}}\right)^2~\frac{1}{4} 
\int\frac{{\rm d}{\bf k}_1{\rm d}{\bf k}_2{\rm d}{\bf k}_3}{(2\pi)^{9}} 
\frac{G^*_{E_1+E_2-E_3}(k_4)-G_{E_1+E_2-E_3}(k_4)}{2\pi i}
\nonumber\\
& & \times \langle{\bf k}_{34}|\hat{T}^{(1)}_U(E_{12})|{\bf k}_{12}\rangle\langle{\bf k}_{13,+}|\left(\hat{T}_U^{(1)}(E'_{13})\right)^\dagger|{\bf k}_{24,+}\rangle\nonumber\\
& & \times
\left|G_{E_1}(k_1)\right|^2 G_{E_2}(k_2)G^*_{\widetilde{E}_2}(k_2) G_{E_3}(k_3)G^*_{\widetilde{E}_3}(k_3)\label{eq:fc}\,,
\end{eqnarray}
where ${\bf k}_4={\bf k}_1+{\bf k}_2-{\bf k}_3$, $E_{12}=E_1+E_2-E_{{\bf k}_1+{\bf k}_2}/2$, $E'_{13}=E_1+\widetilde{E}_3-E_{{\bf k}_1-{\bf k}_3}/2$,
$|{\bf k}_{13,+}\rangle  =  \Bigl(\left|({\bf k}_1+{\bf k}_3)/2\right>+\left| (-{\bf k}_1-{\bf k}_3)/2\right>\Bigr)/\sqrt{2}$,
$|{\bf k}_{24,+}\rangle  =  \Bigl(\left|({\bf k}_2+{\bf k}_4)/2\right>+\left| (-{\bf k}_2-{\bf k}_4)/2\right>\Bigr)/\sqrt{2}$, and 
$|{\bf k}_{12}\rangle,|{\bf k}_{34}\rangle$ as defined after Eq.~(\ref{eq:tmatrix}),
represents the crossed counterpart  of inelastic collision $f_{E_1,E_2,E_3}$, see Fig.~\ref{fig:ladderbb}c). It reduces to {\em two times} the corresponding ladder term, Eq.~(\ref{eq:fsimple}), for $\widetilde{E}_2=\widetilde{E}_3=E_2=E_3$. The factor $2$ originates from the fact that we can reverse the single-particle amplitudes of  the ladder building block, Fig.~\ref{fig:ladderbb}c), also in a different way (with the outgoing dashed arrow pointing to ${\bf r}_1$ instead of ${\bf r}_2$) giving rise to an identical term. The wavevectors ${\bf k}_1, {\bf k}_2$ and ${\bf k}_3$ in Eq.~(\ref{eq:fc}) are associated with Green's functions emitted from (or pointing towards) positions ${\bf r}_1$, ${\bf r}_2$ and ${\bf r}_3$, respectively.

Similarly, there also exist two different possibilities for reversing the ladder building block $g_{E_1,E_2}$, Fig.~\ref{fig:ladderbb}b). Apart from 
$g^{(C)}_{E_1,E_2}$, see Eq.~(\ref{eq:gc}) and Fig.~\ref{fig:crossedbb}b), this gives rise to a new building block, Fig.~\ref{fig:crossedbb}d):
\begin{eqnarray}
h^{(C)}_{E_1,E_2}& = & \left(\frac{4\pi}{\ell_{\rm dis}}\right)^2 
\int\frac{{\rm d}{\bf k}_1{\rm d}{\bf k}_2}{(2\pi)^{6}} G_{E_2}(k_2)G_{\widetilde{E}_2}^*(k_2)\langle{\bf k}_{11}|\hat{T}^{(1)}_U(E_i+E_d)|{\bf k}_{11}\rangle \nonumber\\
& & \times\left|G_{\widetilde{E}_1}(k_1)\right|^2 G_{E_1}(k_1) \label{eq:hc}\,,
\end{eqnarray}
where
$|{\bf k}_{11}\rangle  =  \Bigl(\left|{\bf k}_1\right>+\left| -{\bf k}_1\right>\Bigr)/\sqrt{2}$. The corresponding conjugate diagram, see Fig.~\ref{fig:crossedbb}e), is given by $\left(h^{(C)}_{\widetilde{E}_2,\widetilde{E}_1}\right)^*$. Note that the two colliding particles exhibit opposite wavevectors (${\bf k}_1$ and $-{\bf k}_1$), and therefore the energy of the collision event is fixed to
$\tilde{E}_1+E_1=E_i+E_d$ due to Eq.~(\ref{eq:etilde}).

Each of the above crossed building blocks exhibits an incoming and an outgoing crossed density (defined by the direction of the solid arrow, as mentioned above). Additionally, the two-particle building blocks, Figs.~\ref{fig:crossedbb}b-e), exhibit an incoming ladder density. The latter is given by the solution $I_E({\bf r})$ of the ladder transport equation (\ref{eq:transport2}).

\subsection{Crossed transport equation}

\begin{figure}
\centerline{\includegraphics[width=10cm]{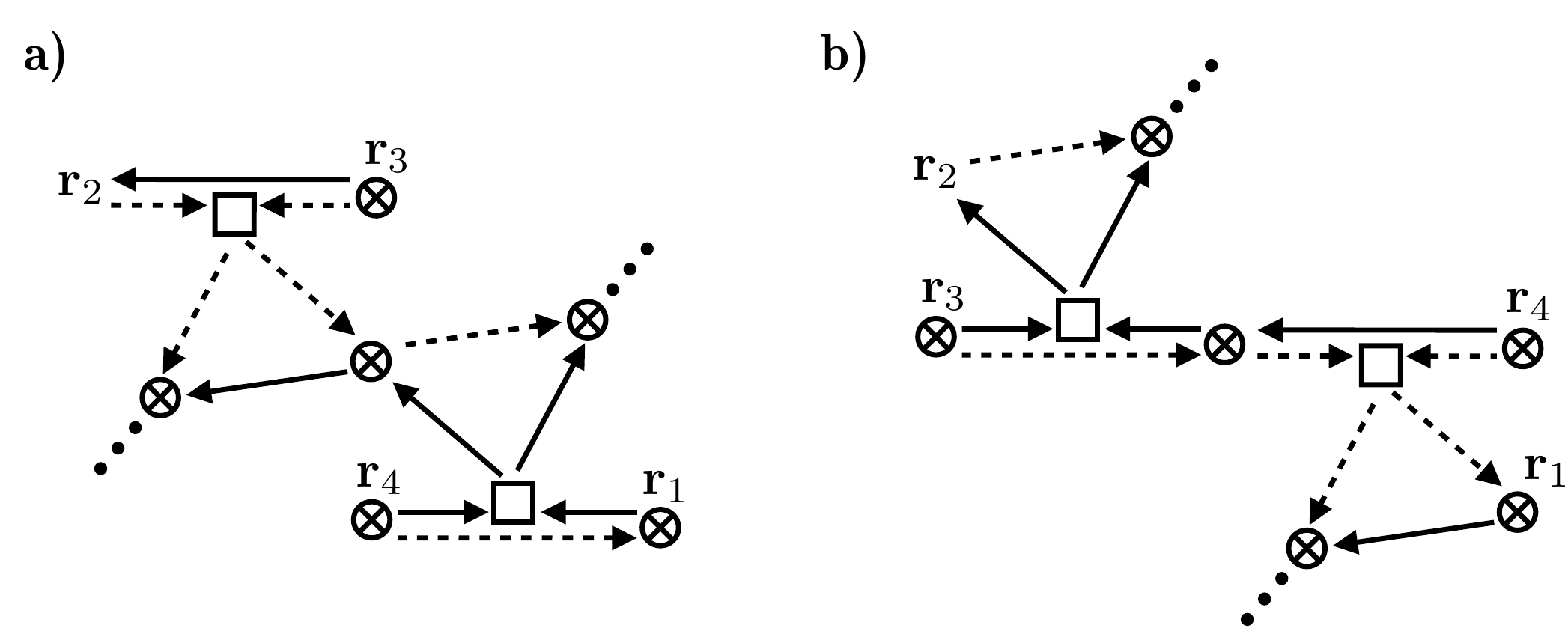}}
\caption{Combinations of crossed building blocks. {\bf a)} When attaching Fig.~\ref{fig:crossedbb}d) to Fig.~\ref{fig:crossedbb}e), the detected atom (solid arrow pointing to ${\bf r}_2$, dashed arrow pointing to ${\bf r}_1$) exhibits no collision with the undetected atoms. Combinations of this type therefore do not contribute to the detection signal, and must be excluded from the crossed transport equation.
{\bf b)} In contrast, the inverse combination, i.e.~attaching Fig.~\ref{fig:crossedbb}e) to Fig.~\ref{fig:crossedbb}d), contributes to the propagation of the crossed density, and must be taken into account in the transport equation.
}
\label{fig:crossedforbidden}
\end{figure}

Propagation of the crossed density can now be described by an integral equation accounting for all possible combinations of the above crossed building blocks (see Fig.~\ref{fig:crossedbb}). An example is displayed in Fig.~\ref{fig:crossedforbidden}a). Here, the outgoing crossed density of the building block shown in Fig.~\ref{fig:crossedbb}d) serves as the incoming crossed density for the building block shown in Fig.~\ref{fig:crossedbb}e). The resulting combination, Fig.~\ref{fig:crossedforbidden}a), exhibits the following remarkable property: if we look at the outgoing arrows (solid arrow pointing to ${\bf r}_2$, dashed arrow pointing to ${\bf r}_1$) corresponding to the detected particle, we see that the detected particle
exhibits no collision with the other particles involved in Fig.~\ref{fig:crossedforbidden}a). The evolution of these undetected particles therefore has no impact on the detected particle and, consequently, as discussed at the end of Sec.~\ref{sec:nparticle}, the process shown in Fig.~\ref{fig:crossedforbidden}a) may be disregarded when calculating the detection signal. The same remains true if -- instead of attaching Fig.~\ref{fig:crossedbb}e) directly to
Fig.~\ref{fig:crossedbb}d) -- an arbitrary sequence of the remaining crossed building blocks, Figs.~\ref{fig:crossedbb}a), b) or c), is inserted in between.
In contrast, for any other combination of building blocks, e.g.~Fig.~\ref{fig:crossedforbidden}b), all involved particles turn out to be connected to each other (through collision events or partial traces), thus contributing to the propagation of the crossed density.

In order to exclude combinations of the former type from the transport equation, we split the crossed density into two parts, i.e.
$C_E({\bf r})=C_E^{(1)}({\bf r})+C_E^{(2)}({\bf r})$. All combinations of the building blocks Figs.~\ref{fig:crossedbb}a-c) and e) are contained in 
$C_E^{(1)}({\bf r})$, and the remaining ones, i.e. those involving Fig.~\ref{fig:crossedbb}d), in $C_E^{(2)}({\bf r})$. According to the rules mentioned above, the building block  Fig.~\ref{fig:crossedbb}e) is excluded from the transport equation for  $C_E^{(2)}({\bf r})$. In total, the transport equations therefore read as follows:
\begin{eqnarray}
C^{(1)}_E({\bf r}) & = & I_0^{(C)}({\bf r})\delta(E-E_i)+\int_{\mathcal V}{\rm d}{\bf r}'P_E^{(C)}({\bf r},{\bf r}')C^{(1)}_E({\bf r}')\nonumber\\
& & +\int_0^\infty{\rm d}E'~\left[g^{(C)}_{E',E} C^{(1)}_E({\bf r})+\int_0^{E_i+E_d}{\rm d}E'' ~f^{(C)}_{E',E'',E} C^{(1)}_{E''}({\bf r})\right]I_{E'}({\bf r})\nonumber\\
& & +\int_0^{E_i+E_d}{\rm d}E'~\left(h^{(C)}_{\widetilde{E},\widetilde{E}'}\right)^* I_{E}({\bf r})C^{(1)}_{E'}({\bf r})\label{eq:c1}\,,
\end{eqnarray}
and
\begin{eqnarray}
C^{(2)}_E({\bf r}) & = & \int_{\mathcal V}{\rm d}{\bf r}'P_E^{(C)}({\bf r},{\bf r}')C^{(2)}_E({\bf r}')\nonumber\\
& & + \int_0^\infty{\rm d}E'\left[g^{(C)}_{E',E} C^{(2)}_E({\bf r})
+\int_0^{E_i+E_d}{\rm d}E'' ~ f^{(C)}_{E',E'',E} C^{(2)}_{E''}({\bf r})\right]I_{E'}({\bf r})\nonumber\\
& & + \int_0^{E_i+E_d}{\rm d}E'~h^{(C)}_{E',E} I_{\widetilde{E}'}({\bf r})\left(C^{(1)}_{E'}({\bf r}'')+C^{(2)}_{E'}({\bf r})\right)\label{eq:c2}\,.
\end{eqnarray}
In Eq.~(\ref{eq:c1}), the incoming crossed density is given by:
\begin{equation}
I^{(C)}_0({\bf r})=\rho_0e^{i{\bf q}\cdot{\bf r}-\left(z_{{\bf r},-\hat{\bf k}_i}+z_{{\bf r},\hat{\bf k}_d}\right)/(2\ell_{\rm dis})}\label{eq:i0c}\,,
\end{equation}
where ${\bf q}={\bf k}_i+{\bf k}_d$, see Eq.~(\ref{eq:q}), where the wave vector of the detected particle is determined by the
energy $E_d$ and the position ${\bf R}$ of the detector (in the far field) as ${\bf k}_d=\sqrt{E_d}{\bf R}/R$, and
$z_{{\bf r},-\hat{\bf k}_i}$ (or $z_{{\bf r},\hat{\bf k}_d}$) corresponds to the distance an incoming (or outgoing) particle travels inside the scattering region, as
defined in Eqs.~(\ref{eq:i0},\ref{eq:jsce}). Finally, the coherently backscattered flux density results as
\begin{equation}
\gamma^{(C)}(\hat{\bf k}_d)=\int_0^\infty{\rm d}E_d~\gamma^{(C)}_{E_d}(\hat{\bf k}_d)\label{eq:jscc}\,,
\end{equation}
with associated spectral density
\begin{eqnarray}
\gamma^{(C)}_{E_d}(\hat{\bf k}_d) & = & 
\int_{\mathcal V} \frac{{\rm d}{\bf r}}{{\mathcal A}\ell_{\rm dis}\rho_0}~\sqrt{\frac{E_d}{E_i}}
e^{-i{\bf q}\cdot{\bf r}-\left(z_{{\bf r},-\hat{\bf k}_i}+z_{{\bf r},\hat{\bf k}_d}\right)/(2\ell_{\rm dis})}
\nonumber\\
& & \times \left[
\left(C^{(1)}_{E_d}({\bf r})+C^{(2)}_{E_d}({\bf r})\right)-\delta(E_d-E_i)I_0^{(C)}({\bf r})\right]\label{eq:jscce}\,,
\end{eqnarray}
where the last term accounts for single scattering. The total average flux measured by a detector placed in direction $\hat{\bf k}_d$, see Eq.~(\ref{eq:gamma}), then corresponds to the sum of the ladder and the crossed component, $\overline{\gamma(\hat{\bf k}_d)}=\gamma^{(L)}(\hat{\bf k}_d)+\gamma^{(C)}(\hat{\bf k}_d)$.

\section{Numerical solutions of the transport equations}
\label{sec:results}

After having developed the general scattering formalism  valid for arbitrary shapes of the interaction potential $U({\bf r})$ and the scattering region $\mathcal V$, we will now focus on the case of a short-range potential $U({\bf r})$ and a slab geometry for $\mathcal V$. As explained in \ref{sec:swave}, the $T$-matrix is then described by a single parameter -- the $s$-wave scattering length $a_s$. The corresponding average distance between (inelastic) collision events is given by:
\begin{equation}
\ell_{\rm int}=\frac{1}{8\pi a_s^2\rho_0}\label{eq:lintswave}\,.
\end{equation}
In the following, we measure the interaction strength in terms of the ratio between $\ell_{\rm dis}$ and $\ell_{\rm int}$: 
\begin{equation}
\alpha=\frac{\ell_{\rm dis}}{\ell_{\rm int}}=8\pi a_s^2\ell_{\rm dis}\rho_0\label{eq:alpha}\,,
\end{equation}
which, as explained in Sec.~\ref{sec:ladder}, should fulfill the condition $\alpha\ll 1$, and, due to Eq.~(\ref{eq:lintswave}), is proportional to $a_s^2$. Indeed, we see from Eqs.~(\ref{eq:gswave},\ref{eq:fswave})
that the ladder collision terms $g$ and $f$ both depend on $\alpha$. The terms proportional to $a_s$ in $g$ drop out as a consequence of flux conservation, see Eq.~(\ref{eq:particlecons}). The same is not true for the crossed collision terms $g^{(C)}$ and $h^{(C)}$, see Eqs.~(\ref{eq:gcswave},\ref{eq:hcswave}), which depend on a second parameter proportional to $a_s$:
\begin{equation}
\beta=\frac{8\pi a_s\ell_{\rm dis}\rho_0}{\sqrt{E_i}}=\frac{\alpha}{\sqrt{E_i}a_s}\label{eq:beta}\,.
\end{equation}
Since $\sqrt{E_i}a_s\ll 1$ for $s$-wave scattering, it follows that $\beta\gg\alpha$.  The parameter $\beta$ can also be expressed in terms of the
healing length $\xi=(8\pi\rho_0 a_s)^{-1/2}$ \cite{pethick08}, i.e.~$\beta=\ell_{\rm dis}/(\sqrt{E_i}\xi^2)$, or in terms of the interaction parameter $g=8\pi a_s$ appearing in the Gross-Pitaevskii equation, i.e. $\beta=g\rho_0\ell_{\rm dis}/\sqrt{E_i}$. The Gross-Pitaevskii equation is valid in the limit
$a_s\to 0$ and $\rho_0\to\infty$ \cite{lieb05} such that $a_s\rho_0={\rm const}$. Since $\alpha\to 0$ in this limit, our previously derived transport equations for nonlinear coherent backscattering based on the Gross-Pitaevskii equation \cite{hartung08,wellens09b,wellens09c} must be recovered from Eqs.~(\ref{eq:transport2},\ref{eq:c1},\ref{eq:c2})  for $\alpha=0$, as it is indeed the case if we insert the $s$-wave expressions, Eqs.~(\ref{eq:gswave}-\ref{eq:fcswave}), evaluated at $\alpha=0$.

Concerning the geometry of the scattering medium, we choose a slab confined between two planes, $z=0$ and $z=L$, respectively, with perpendicular incident wavevector, i.e. ${\bf k}_i=(0,0,k_i)$. The thickness of the slab in units of the disorder mean free path defines its optical thickness $b=L/\ell_{\rm dis}$. The slab geometry is very convenient from a numerical point of view, since the integration over $x$ and $y$ can be performed analytically in Eqs.~(\ref{eq:transport2},\ref{eq:c1},\ref{eq:c2}), such that the resulting transport equations only depend on $z$ \cite{mark88}. Moreover, due to rotational symmetry around the $z$-axis, the backscattered flux $g(\hat{\bf k}_d)=g(\theta)$ depends only on the backscattering angle $\theta$ defined by $\hat{\bf k}_i\cdot \hat{\bf k}_d=-\cos\theta$, and the distances appearing in Eqs.~(\ref{eq:i0},\ref{eq:jsce},\ref{eq:i0c},\ref{eq:jscce}) simplify to $z_{{\bf r},-\hat{\bf k}_i}=z$ and $z_{{\bf r},\hat{\bf k}_d}=z/\cos\theta$, respectively. Finally, the integration over
the scattering volume $\mathcal V$ in Eqs.~(\ref{eq:jsce},\ref{eq:jscce}) reduces to $\int_{\mathcal V} {\rm d}{\bf r}/{\mathcal A}\to \int_0^L{\rm d}z$. The one-dimensional versions of the transport equations (\ref{eq:transport2},\ref{eq:c1},\ref{eq:c2}) can now be solved numerically, e.g. by iteration.

\subsection{Density inside the slab}

\begin{figure}
\centerline{\includegraphics[width=8cm]{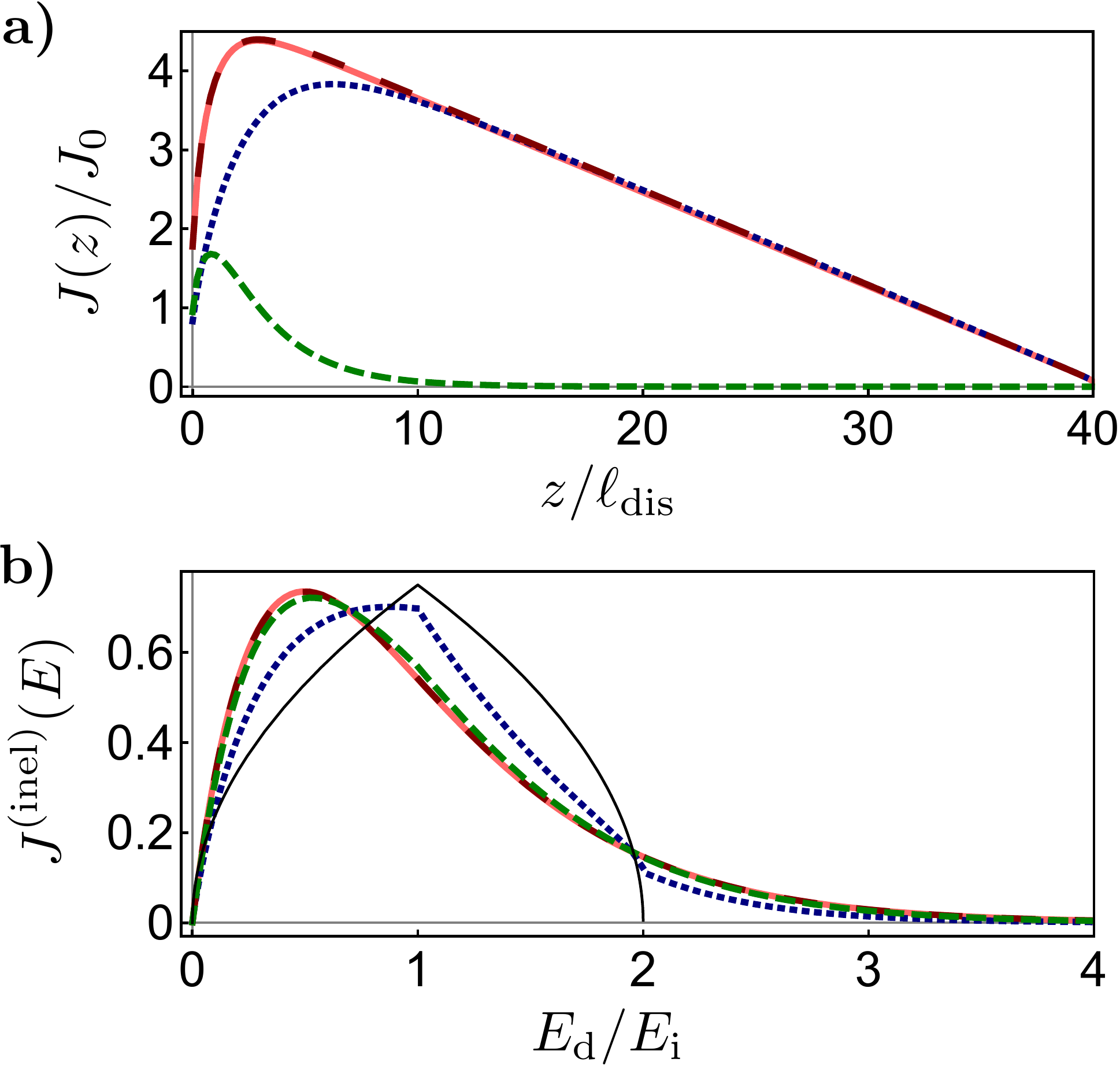}}
\caption{(Color online) {\bf a)} Different components of the average flux density $J(z)$, plotted as a function of position $z$ in the slab, for weak interaction $\alpha=1/100$, and thickness $b=40$. The linear flux density (red solid line) coincides with the total flux density for the case of many particles (black long-dashed). The latter splits into an elastic (green dashed) and inelastic component (blue dotted). In spite of the weak interaction, the transport is dominated by inelastically scattered particles, especially deep inside the slab. {\bf b)} Normalized energy distribution $J_E^{({\rm inel})}(z)$ of inelastically scattered atoms for different positions $z=0$ (blue dotted), $z=L/4$ (green dashed) and $z=L$ (black long-dashed) in the slab, and otherwise the same parameters as in a). The thin black line displays $\sqrt{E}f_{E_{\rm i},E_{\rm i},E}/(-\sqrt{E_{\rm i}}g_{E_{\rm i},E_{\rm i}})$ (see Eq.~(\ref{eq:particlecons}) for the normalization) according to Eq.~(\ref{eq:fswave}), i.e.~the distribution after a single inelastic scattering event. The kink of this distribution is recovered at the beginning of the slab (i.e., for $z=0$). Deep inside the slab (i.e., for $z=L/4$ and $z=L$), the spectrum collapses onto a thermal Maxwell-Boltzmann distribution with average energy $E_{{\rm av}}=E_{\rm i}$ (red solid).}
\label{fig:ladderresult}
\end{figure}

Fig.~\ref{fig:ladderresult}a) shows the resulting flux density $J(z)=\int_0^\infty{\rm d}E~\sqrt{E}I_E(z)$, see Eq.~(\ref{eq:jk}), as a function of the position
$z$ inside the slab, for weak interactions $\alpha=1/100$, cf. Eq.~(\ref{eq:alpha}), and optical thickness $b=40$. 
As already proven after Eq.~(\ref{eq:jtot}), the total flux $J(z)$  (black long-dashed line) equals the linear flux (red solid) as obtained from Eq.~(\ref{eq:transport2}) with $\alpha=0$. In contrast to the linear case, however, the flux $J(z)$ splits into an elastic (green dashed) and an inelastic component (blue dotted), defined by $J_E(z)=
J_E^{\rm (el)}(z)\delta(E-E_i)+J_E^{\rm (inel)}(z)$. We see that, in spite of the weakness of the interaction ($\alpha=1/100$), the inelastic component dominates, especially deep inside the slab. This can be explained by the large number ($\approx b^2$) of scattering events required to traverse a slab with thickness $b$. The expected number of two-body collision events thus results as approximately $\alpha b^2=16$.
Let us note that the inelastic component of the flux is associated with a non-condensed fraction of atoms, since an $N$-fold product of a single-particle state (as required from the formal definition of a condensate via the stationary one-particle density matrix \cite{lieb05}) with fixed total energy implies fixed energies also for the individual particles. 

The normalized energy distribution $J_E^{(\rm inel)}(z)$ of the inelastic component is shown in Fig.~\ref{fig:ladderresult}b), for different positions $z$ inside the slab. We see that, far inside the slab, i.e. at $z=10\ell_{\rm dis}$ (green dashed) and $z=40\ell_{\rm dis}$ (black long-dashed), the energy distribution approaches a Maxwell-Boltzmann distribution
$J^{(MB)}_E/J=4 E\exp(-2E/E_i)/E_i^2$ (red solid), see Eq.~(\ref{eq:mb}). Thereby, we confirm the analytical result derived in Sec.~\ref{sec:thermalization} for an infinite slab. In contrast, at the beginning of the slab (blue dotted), the distribution is not yet thermalized, and lies  between the Maxwell-Boltzmann distribution and the distribution $\sqrt{E}f_{E_{\rm i},E_{\rm i},E}/(-\sqrt{E_{\rm i}}g_{E_{\rm i},E_{\rm i}})$ obtained after a single inelastic collision event (thin black line) and normalized according to Eq.~(\ref{eq:particlecons}).

\subsection{Backscattered flux outside the slab}

\begin{figure}
\centerline{\includegraphics[width=8cm]{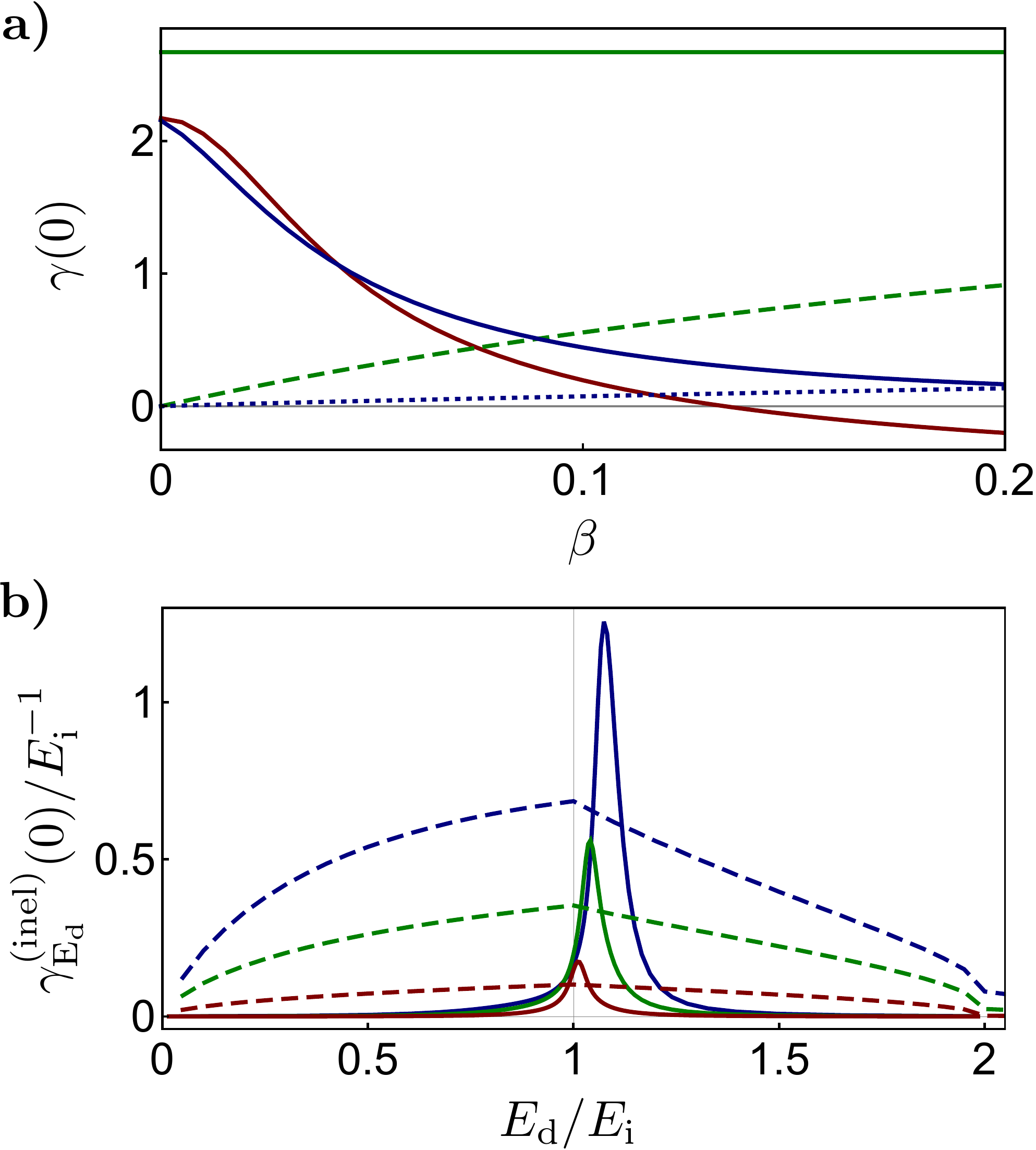}}
\caption{(Color online) {\bf a)} Background and interference contributions $\gamma^{(L)}(0)$ and $\gamma^{(C)}(0)$ to the scattered flux in exact backscattering direction ($\theta=0$) for a slab with thickness $b=10$ and $\ell_{\rm dis}\sqrt{E_i}=10$ as a function of the crossed collision strength $\beta$, see Eq.~(\ref{eq:beta}). The Gross-Pitaevskii equation ($\alpha=0$, red line) predicts a crossover from constructive ($\gamma^{(C)}(0)>0$) to destructive interference ($\gamma^{(C)}(0)<0$) at $\beta\simeq 0.13$. In presence of inelastic collisions ($\alpha=\beta/10$, blue solid line), the decrease of $\gamma^{(C)}(0)$ is initially faster, but for $\beta>0.04$ slower than in the case $\alpha=0$, due to the steadily rising inelastic component $\gamma^{(C,{\rm inel})}(0)$ (blue dotted line).  For comparison, the total background contribution $\gamma^{(L)}(0)$ (green solid), which is independent of $\alpha$ and $\beta$, and its inelastic component $\gamma^{(L,{\rm inel})}(0)$ (for $\alpha=\beta/10$, green dashed) are also shown.
 {\bf b)} Spectral distributions $\gamma^{(C,{\rm inel})}_{E_d}(0)$ (solid lines) and $\gamma^{(L,{\rm inel})}_{E_d}(0)$ (dashed lines) for the same parameters as in a) and
 $\beta=0.02$ (red), 0.08 (green) and 0.2 (blue). In a narrow spectral region around $E_d\simeq E_i$ (but slightly shifted with respect to $E_i$), the interference contribution $\gamma^{(C,{\rm inel})}_E(0)$ exceeds the background $\gamma^{(L,{\rm inel})}_E(0)$, corresponding to a coherent backscattering enhancement factor larger than two.}
\label{fig:crossedresult}
\end{figure}

Fig.~\ref{fig:crossedresult}a) shows the background and interference contributions to the backscattered flux in exact backscattering direction $\theta=0$ for $\ell_{\rm dis}\sqrt{E_i}=10$. Since, in general, the backscattered flux is dominated by scattering paths which do not penetrate very deeply into the slab, we restrict ourselves to the case of a moderate slab thickness $b=10$ (as compared to $b=40$ in Fig.~\ref{fig:ladderresult}). As already known from previous work on nonlinear coherent backscattering in the purely elastic Gross-Pitaevskii limit \cite{hartung08,wellens09b,wellens09c}, 
the backscattered flux $\gamma^{(C)}(0)$ for $\alpha=0$ (red line) exhibits a transition from constructive to destructive interference 
for increasing $\beta$. This transition can be explained by the fact that the nonlinearity effectively introduces dephasing between two reversed scattering paths \cite{hartung08,wellens09b,wellens09c}. For larger $\beta$, the phase difference accumulated along the shortest scattering paths, i.e. those exhibiting only a few, but at least two scattering events (since, as mentioned above, single scattering does not contribute), may exceed $\pi/2$, and thus lead to destructive interference. This behavior changes if a finite amount of inelastic scattering, $\alpha=\beta/10$ corresponding to $a_s\sqrt{E_i}=1/10$, see Eq.~(\ref{eq:beta}), is taken into account (blue line): At first, the interference drops faster than for $\alpha=0$, since inelastic scattering changes the frequency and thus leads to an additional dephasing mechanism. At larger values $\beta$, however, the decrease of the backscattered flux is slowed down as compared to the purely elastic case, such that a transition to destructive interference is not observed in Fig.~\ref{fig:crossedresult}a).

This behaviour can be explained by examining the spectral distribution $\gamma_{E_d}^{(C,{\rm inel})}$ of the inelastic interference component as a function of the detected frequency $E_d$ in Fig.~\ref{fig:crossedresult}b). Since the amount of inelastic scattering is governed by $\alpha=\beta/10$, both  $\gamma_{E_d}^{(L,{\rm inel})}(0)$ (dashed lines) and  $\gamma_{E_d}^{(C,{\rm inel})}(0)$ (solid lines) increase as a function of $\beta$ (green, blue and red lines corresponding to $\beta=0.02$, $0.08$ and $0.2$, respectively.) The interference spectra $\gamma_{E_d}^{(C,{\rm inel})}(0)$ exhibit narrow peaks close to the initial energy, $E_d\simeq E_i$.
The relative width of these peaks approximately equals $\Delta E_d/E_i\simeq 1/(\ell_{\rm dis}\sqrt{E_d})=1/10$, which can be understood as a consequence of frequency-induced dephasing given by Eq.~(\ref{eq:avintc}) with $E\neq \widetilde{E}=E_i+E_d-E$. Less intuitive is the
fact that the maxima of the peaks are slightly shifted with respect to $E_i$, for which, at present, we are lacking a simple explanation.
Please note, however, that the interference peaks exceed the background (most remarkably for larger values of $\beta$), corresponding to a coherent backscattering enhancement factor of the inelastic flux contribution $\eta^{({\rm inel})}_{E_d}=\left(\gamma_{E_d}^{(C,{\rm inel})}(0)+
\gamma_{E_d}^{(L,{\rm inel})}(0)\right)/\gamma_{E_d}^{(L,{\rm inel})}(0)>2$ within a narrow spectral window around the energy $E_d$ where the crossed contribution is maximal. This enhancement is a consequence of the many-wave interference character of nonlinear coherent backscattering \cite{wellens06,wellens06b} resulting from the fact that, as discussed in Sec.~\ref{sec:crossedbb}, there are several ways of reversing the scattering paths when constructing crossed from ladder diagrams. The number of these possibilities increases with increasing number of inelastic scattering events. Although, due to the small width of these peaks, the total inelastic component $\gamma^{(C,{\rm inel})}(0)$ (integrated over $E_d$) turns out to be smaller than the background $\gamma^{(L,{\rm inel})}(0)$ [cf.~the dashed blue and dotted green lines in Fig.~\ref{fig:crossedresult}a)], this many-wave interference effect contributes to the above observed slowing down of the decrease of the backscattered flux. 

\section{Conclusion}
\label{sec:concl}

Within this paper, we have derived a microscopic $N$-body scattering theory for interacting particles in a weak disorder potential in three dimensions. We have applied this diagrammatic theory to a stationary scattering scenario for an asymptotically non-interacting quasi-plane matter wave incident on a three-dimensional slab, with the disorder potential and inter-particle collisions confined to the slab region, and hereby verified the viability of our theory to address, on the one hand, very fundamental but, on the other hand, very timely questions of quantum transport for interacting particles in random environments. In a clear and precise manner, we demonstrated how one  can bridge the gap between strictly unitary many-body evolution and its implications on the mesoscopic level governed by a transport equation similar to the nonlinear Boltzmann transport equation. Furthermore, we have determined the coherent corrections due to the wave nature of the particles leading to the effect of coherent backscattering. We have demonstrated
that inelastic scattering slows down the decrease of the coherent backscattering peak as compared to the purely elastic case described by the Gross-Pitaevskii equation.

Let us briefly summarize the basic assumptions of our theory: first, we assume an optically thick scattering medium ($b=L/\ell_{\rm dis}\gg 1$) allowing for multiple scattering in a weak disorder potential, where the mean free path is much larger than the wavelength of the incoming particles ($\ell_{\rm dis}\sqrt{E_i}\gg 1$). Second, the interaction strength as, respectively, quantified by the parameters $\alpha$ and $\beta$ for ladder and crossed collisions, see Eqs.~(\ref{eq:alpha},\ref{eq:beta}) for the case of $s$-wave scattering, should fulfill the condition $\alpha,\beta\ll 1$, such that atom-atom collisions occur less frequently than disorder scattering events. With view at future work, we expect that the condition $\beta\ll 1$ can be dropped by summing the corresponding diagrams, see Fig.~\ref{fig:crossedbb}b,d,e), without intermediate disorder scattering \cite{wellens09b}. If
$\beta^2\simeq \ell_{\rm dis}\sqrt{E_i}$, another type of collision process -- corresponding to scattering induced by the fluctuating background density -- sets in which is described  by diagrams similar to our ladder diagrams \cite{schwiete13}. Furthermore, relaxing the contact approximation for the collision terms, Eqs.~(\ref{eq:gcontact},\ref{eq:fcontact}), will allow to determine the effect of attractive or repulsive interactions onto the spatial atomic density profile. It will hence be a worthwhile task to extend our theory to stronger interactions, although we surely expect to encounter certain limits (e.g.,~the regime of superfluidity) where other methods will be required.

Concerning an experimental verification of our results, the application to a stationary scattering setup with matter waves constitutes, on the one hand, a very timely scenario, as, e.g.,~the developments of atom lasers and matter wave interferometers on atom chips \cite{guerin06,couvert08,fortagh07} rapidly progress. On the other hand, many years of expertise have been gathered within the field of wave-packet spreading upon releasing the condensate from a trap into a new environment, where, e.g.,~the first experiments on coherent backscattering~of (non-interacting) matter waves have been reported recently \cite{labeyrie12,jendrzejewski12}. Consequently, an extension of our theory to time-dependent scenarios based on recent progress in this field \cite{schwiete13b,schwiete10,cherroret11} presents a significant and feasible task. Similarly, also a finite correlation length of the disorder potential can be taken into account, see \cite{kuhn05,kuhn07} for the non-interacting case. 

In conclusion, we are confident, that our present theory and the rather straightforward extensions discussed above will substantially foster a more complete understanding of quantum transport under the interplay of disorder and inter-particle interaction and can contribute to a unifying picture from microscopic to macroscopic scales.

\section*{Acknowledgements}

We thank Nicolas Cherroret, Pierre Lugan, Cord A. M\"uller and Peter Schlagheck for fruitful discussions.
We acknowledge partial support by DFG research unit FG760. T. G. acknowledges funding through DFG Grant No. BU1337/8-1.

\appendix

\section{Factorization of the transition amplitude}
\label{sec:factorization}

\begin{figure}
\centerline{\includegraphics[width=4cm]{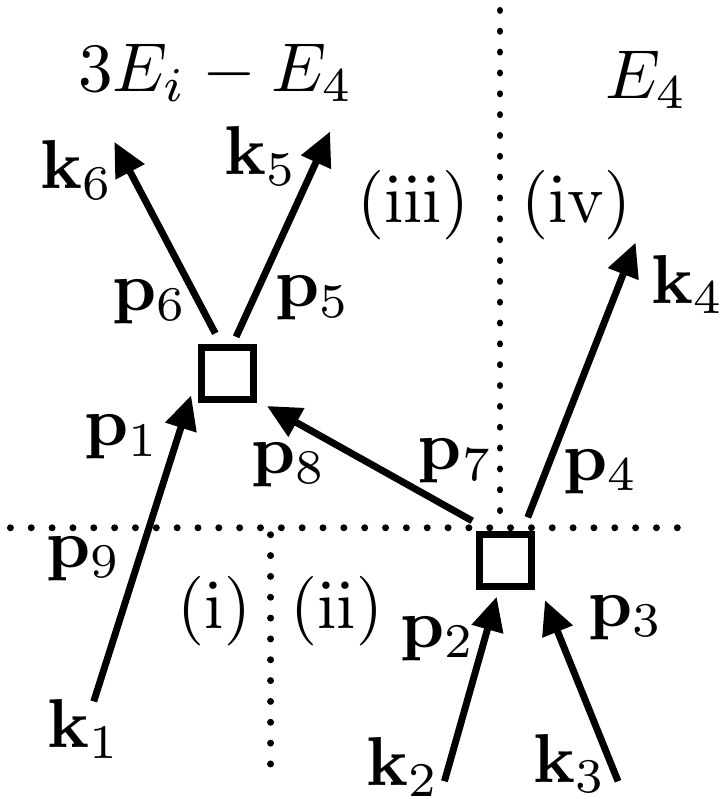}}
\caption{The dashed lines split the diagram of  Fig.~\ref{fig:amplitudes} into 4 subdiagrams (i), (ii), (iii), and (iv). Note that the subdiagrams (i) and (ii) -- and likewise (iii) and (iv) -- are not connected to each other. This allows us to factorize the 3-particle diagram into 1- and 2-particle diagrams.}
\label{fig:amplitudes2}
\end{figure}

In this appendix, we show how an arbitrary $N$-particle scattering diagram can be factorized into single-particle propagators and two-body collisions. We first look at the example diagram shown in Fig.~\ref{fig:amplitudes}. As shown in Fig.~\ref{fig:amplitudes2}, this diagram can be split into four independent subdiagrams. 
The two subdiagrams connected to the initial state -- (i) and (ii) in Fig.~\ref{fig:amplitudes2} -- correspond to M\o ller operators, and the remaining ones  -- (iii) and (iv) --  to Green's operators. This gives rise to the following matrix elements:
\begin{eqnarray}
\Omega^{\rm (i)}_+(E) & = & \langle{\bf p}_9|\hat{\Omega}^{(V)}_+(E)|{\bf k}_1\rangle\,,\\
\Omega^{\rm (ii)}_+(E) & = & \frac{1}{2}\int\frac{{\rm d}{\bf p}_2{\rm d}{\bf p}_3}{(2\pi)^6}
 \langle {\bf p}_4,{\bf p}_7|\hat{T}_U(E)|{\bf p}_2,{\bf p}_3\rangle\langle{\bf p}_2,{\bf p}_3|\hat{\Omega}^{(V)}(E)|{\bf k}_2,{\bf k}_3\rangle
 \label{eq:omII}\,,\\
G^{\rm (iii)}(E) & = & \frac{1}{4}\int\frac{{\rm d}{\bf p}_1{\rm d}{\bf p}_5{\rm d}{\bf p}_6{\rm d}
{\bf p}_8}{(2\pi)^{12}}
\langle{\bf k}_5,{\bf k}_6|\hat{G}_V(E)|{\bf p}_5,{\bf p}_6\rangle\label{eq:gIII}\\
& & \times \langle {\bf p}_5,{\bf p}_6|\hat{T}_U(E)|{\bf p}_1,{\bf p}_8\rangle\langle{\bf p}_1,{\bf p}_8|\hat{G}_V(E)|{\bf p}_7,{\bf p}_9\rangle\nonumber\,,\\
G^{\rm (iv)}(E) & = & \langle{\bf k}_4|\hat{G}_V(E)|{\bf p}_4\rangle\,.
\end{eqnarray}
Note that the diagrams (i) and (ii) are not connected to each other in Fig.~\ref{fig:amplitudes2}. The corresponding M\o ller operators can therefore be factorized as in Eq.~(\ref{eq:factorizeom}). Likewise, the Green's functions corresponding to (iii) and (iv) are factorized according to Eq.~(\ref{eq:factorize}). The prefactors $1/2$ and $1/4$ in Eqs.~(\ref{eq:omII},\ref{eq:gIII}) originate from symmetrization in the two-particle subspace (e.g. the states $|{\bf p}_2,{\bf p}_3\rangle$ and $|{\bf p}_3,{\bf p}_2\rangle$ are identical and therefore must not be summed over twice). It turns out that these factors are compensated, however, by the two possibilities to associate the initial and final single-particle states with each other in Eqs.~(\ref{eq:factorizeom},\ref{eq:factorize}).

The total transition amplitude results as:
\begin{eqnarray}
& & \langle{\bf k}_4,{\bf k}_5,{\bf k}_6|\hat{\Omega}_+^{\rm (fig.\ref{fig:amplitudes})}(3E_i)|{\bf k}_1,{\bf k}_2,{\bf k}_3\rangle  = 
\frac{1}{2}\int\frac{{\rm d}{\bf p}_4{\rm d}{\bf p}_7{\rm d}{\bf p}_9}{(2\pi)^9} \Omega^{\rm (i)}_+(E_i)\Omega^{\rm (ii)}_+(2E_i)\nonumber\\
& &\times \int_{-\infty}^\infty \frac{{\rm d}E_4}{(-2\pi i)}G^{\rm (iii)}(3E_i-E_4)G^{\rm (iv)}(E_4)\,.
\end{eqnarray}
Now, we again apply Eqs.~(\ref{eq:factorizeom},\ref{eq:factorize}) to factorize the two-particle M\o ller  and Green's operators on the right-hand side of Eqs.~(\ref{eq:omII},\ref{eq:gIII}) into 
single-particle operators. In this way, we recover most of the terms in Eq.~(\ref{eq:smatrix3}). The only ones which appear to differ from Eq.~(\ref{eq:smatrix3}) are those associated to ${\bf k}_1$, ${\bf p}_1$, ${\bf p}_7$ and ${\bf p}_8$, which we reformulate as follows:
\begin{eqnarray}
& & \frac{1}{2}\int\frac{{\rm d}{\bf p}_9}{(2\pi)^3} \langle {\bf p}_1,{\bf p}_8|\hat{G}_V(3E_i-E_4)|{\bf p}_7,{\bf p}_9\rangle\langle{\bf p}_9|\hat{\Omega}^{(V)}_+(E_i)|{\bf k}_1\rangle\nonumber\\
& & = \int_{-\infty}^\infty \frac{{\rm d}E_1}{(-2\pi i)} \langle{\bf p}_1|\hat{G}_V(E_1)\hat{\Omega}_+^{(V)}(E_i)|{\bf k}_1\rangle 
G(-E_1)\label{eq:appfactorize1}\,,
\end{eqnarray}
where we again applied Eq.~(\ref{eq:factorize}), used the completeness relation $\int{\rm d}{\bf p}_9 |{\bf p}_9\rangle\langle{\bf p}_9|=(2\pi)^3$, and defined:
\begin{equation}
G(-E_1)=\langle{\bf p}_8|\hat{G}_V(3E_i-E_4-E_1)|{\bf p}_7\rangle\label{eq:gminus}\,.
\end{equation}
Note that $G(-E_1)$ is a complex analytic function of $E_1$ with poles only in the upper half of the complex plane. This, again, is due to the fact that $\hat{G}_V(E)$ as a function of $E$ exhibits poles only in the lower half, whereas $E_1$ enters with negative sign in the right hand side of Eq.~(\ref{eq:gminus}).
We now reformulate some terms in Eq.~(\ref{eq:appfactorize1}) as follows:
\begin{eqnarray}
& & \hat{G}_V(E_1)\hat{\Omega}_+^{(V)}(E_i)|{\bf k}_1\rangle= \left({\mathbbm 1}+\hat{G}_V(E_1)\hat{V}\right)\hat{G}_0(E_1)|{\bf k}_1\rangle+\hat{G}_V(E_1)\hat{G}_V(E_i)\hat{V}|{\bf k}_1\rangle\nonumber\\
& & = \frac{1}{E_1-E_i+i\epsilon}\left[{\mathbbm 1}+\left(\hat{G}_V(E_1)+\hat{G}_V(E_i)-\hat{G}_V(E_1)\right)\hat{V}\right]|{\bf k}_1\rangle
\nonumber \\
& & = \frac{1}{E_1-E_i+i\epsilon} \hat{\Omega}_+^{(V)}(E_i)|{\bf k}_1\rangle\label{eq:appfactorize2}\,,
\end{eqnarray}
where we used Eq.~(\ref{eq:omegav}), the alternative but equivalent expression $\hat{G}_V(E) = \hat{G}_0(E)+\hat{G}_V(E)\hat{V}\hat{G}_0(E)$ with respect to Eq.~(\ref{eq:lippmannschwingergv}), and the identity
\begin{equation}
\hat{G}_V(E_1)\hat{G}_V(E_i) = \frac{1}{E_1-E_i+i\epsilon}\left(\hat{G}_V(E_i)-\hat{G}_V(E_1)\right)\label{eq:gg}\,,
\end{equation}
resulting from $\frac{1}{ab}=\frac{1}{b-a}\left(\frac{1}{a}-\frac{1}{b}\right)$ (where we set the imaginary part in the denominator of $\hat{G}_V(E_1)$, see Eq.~(\ref{eq:gv}), equal to $2\epsilon$ instead of $\epsilon$, and used the fact that $\hat{G}_V(E_1)$ and $\hat{G}_V(E_i)$ have the same set of eigenvectors). After inserting Eq.~(\ref{eq:appfactorize2}) into Eq.~(\ref{eq:appfactorize1}), we perform the integral over $E_1$ by closing the integration contour in the lower half of the complex plane. (Remember that $G(-E_1)$ has no poles in the lower half!) Thereby, the energy $E_1$ is set to $E_i$, and we finally recover the missing terms in Eq.~(\ref{eq:smatrix3}):
\begin{eqnarray}
& & \int_{-\infty}^\infty \frac{{\rm d}E_1}{(-2\pi i)} \langle{\bf p}_1|\hat{G}_V(E_1)\hat{\Omega}_+^{(V)}(E_i)|{\bf k}_1\rangle 
G(-E_1) \nonumber\\
& & = \langle{\bf p}_1|\hat{\Omega}_+^{(V)}(E_i)|{\bf k}_1\rangle 
G(-E_i)\label{eq:omgmerge}\,.
\end{eqnarray}
The above procedure can be generalized to an arbitrary many-particle scattering diagram: We first divide the whole diagram into independent subdiagrams. Then, some of these subdiagrams turn out to be connected to each other by single-particle propagators. In the above example, Fig.~\ref{fig:amplitudes2}, this is the case for the subdiagram (i) and (iii), which are connected by the single-atom propagators from ${\bf k}_1$ to ${\bf p}_9$ with energy $E_i$ and from ${\bf p}_9$ to ${\bf p}_1$ with energy $E_1$. We have to show that these propagators merge into a single propagator (from ${\bf k}_1$ to
${\bf p}_1$ with energy $E_i$). For the case that one of the propagators is connected to the initial state (and thus corresponds to a M\o ller operator), the corresponding general identity is given by Eq.~(\ref{eq:omgmerge}). If both propagators correspond to Green's operators, e.g. $\hat{G}_V(E_1)$ and $\hat{G}_V(E_2)$ below, the required identity is proven as follows:
\begin{eqnarray}
& & \int_{-\infty}^\infty \frac{{\rm d}E_1{\rm d}E_2}{(-2\pi i)^2} \hat{G}_V(E_1)\hat{G}_V(E_2)G^{(1)}(-E_1)G^{(2)}(-E_2)\nonumber\\
& & =  \int_{-\infty}^\infty \frac{{\rm d}E_1{\rm d}E_2}{(-2\pi i)^2}\frac{1}{E_1-E_2+i\epsilon}\left(\hat{G}_V(E_2)-\hat{G}_V(E_1)\right) G^{(1)}(-E_1)G^{(2)}(-E_2)
\nonumber\\
& & = \int_{-\infty}^\infty \frac{{\rm d}E_2}{(-2\pi i)} \hat{G}_V(E_2) G^{(1)}(-E_2)G^{(2)}(-E_2)\label{eq:ggmerge}\,,
\end{eqnarray}
where we again used Eq.~(\ref{eq:gg}) and the fact that $G^{(1)}(-E_1)$ and $G^{(2)}(-E_2)$ (which correspond to arbitrary other subdiagrams where the energy $E_1$ or $E_2$ enters with negative sign) exhibit no pole in the lower half of the complex plane. Note that the term with $\hat{G}_V(E_1)$ in the second line of Eq.~(\ref{eq:ggmerge}) vanishes after integrating over $E_2$, since no pole remains in the lower half. In total, the concatenation of two Green's operators, i.e. $\hat{G}_V(E_1)\hat{G}_V(E_2)$ on the left-hand-side of Eq.~(\ref{eq:ggmerge}), reduces to a single Green's operator, i.e. $\hat{G}_V(E_2)$ on the left-hand-side, whereas the energy $E_1$ is set equal to $E_2$.

\section{Inelastic and elastic example diagrams}
\label{sec:examplediag}

The inelastic diagram shown in Fig.~\ref{fig:trace1}a) gives the following contribution to the flux density:
\begin{eqnarray}
& & {\bf J}^{\rm (fig.\ref{fig:trace1}a)}({\bf r})  =  \left(\frac{1}{2}\right)^3 2^2 \int\frac{{\rm d}{\bf k}{\rm d}{\bf k}_1{\rm d}{\bf k}_2{\rm d}{\bf k}_3{\rm d}{\bf k}'_1{\rm d}{\bf k}_2' {\rm d}{\bf k}'_3{\rm d}{\bf p}_1{\rm d}{\bf p}_2{\rm d}{\bf p}_3{\rm d}{\bf p}_4{\rm d}{\bf p}'_1{\rm d}{\bf p}_2'{\rm d}{\bf p}_3' {\rm d}{\bf p}'_4}{(2\pi)^{45}}
\nonumber\\
&  &\times \int_{-\infty}^\infty \frac{{\rm d}E{\rm d}E'}{|2\pi i|^2} w^*({\bf k}'_1)w^*({\bf k}'_2)w({\bf k}_1)w({\bf k}_2)\langle{\bf k}'_1|\left(\hat{\Omega}^{(V)}(E_i)\right)^\dagger|{\bf p}'_1\rangle
\langle{\bf k}'_2|\left(\hat{\Omega}^{(V)}(E_i)\right)^\dagger|{\bf p}'_2\rangle\nonumber\\
& & \times \langle{\bf p}'_1,{\bf p}'_2|\left(\hat{T}_U(2E_i)\right)^\dagger|{\bf p}'_3,{\bf p}'_4\rangle
\langle {\bf p}'_3|\left(\hat{G}_V(2E_i-E')\right)^\dagger|{\bf k}'_3\rangle  
\langle{\bf k}'_3|\hat{\bf J}({\bf r})|{\bf k}_3\rangle\nonumber\\
& & \times\langle {\bf p}'_4|\left(\hat{G}_V(E')\right)^\dagger|{\bf k}\rangle\langle {\bf k}|\hat{G}_V(E)|{\bf p}_4\rangle
\langle {\bf k}_3|\hat{G}_V(2E_i-E)|{\bf p}_3\rangle \nonumber\\
& & \times  \langle{\bf p}_3,{\bf p}_4|\hat{T}_U(2E_i)|{\bf p}_1,{\bf p}_2\rangle 
\langle{\bf p}_1|\hat{\Omega}^{(V)}(E_i)|{\bf k}_1\rangle\langle{\bf p}_2|\hat{\Omega}^{(V)}(E_i)|{\bf k}_2\rangle \label{eq:figtrace1a}\,.
\end{eqnarray}
Here, we used the following labels for the wave vectors: the incoming solid arrows are called ${\bf k}_1$ and ${\bf k}_2$, whereas the detected and the traced out solid arrows are given by ${\bf k}_3$ and ${\bf k}$, respectively. The intermediate solid arrows before and after the interaction event are labeled by ${\bf p}_1$ and ${\bf p}_2$ and by ${\bf p}_3$ and ${\bf p}_4$, respectively. The same notation holds for the dashed arrows, which are, however, denoted by an additional prime.

The trace formula, Eq.~(\ref{eq:trace3}), can now be applied as follows: (i) replace $\langle {\bf p}'_4|\left(\hat{G}_V(E')\right)^\dagger|{\bf k}\rangle\langle {\bf k}|\hat{G}_V(E)|{\bf p}_4\rangle$ by $\langle {\bf p}'_4|\left[\left(\hat{G}_V(E)\right)^\dagger-\hat{G}_V(E)\right]|{\bf p}_4\rangle$, (ii) delete the integrals $\int{\rm d}{\bf k}/(2\pi)^3$ and $\int{\rm d}E'/(-2\pi i)$, and (iii) replace $E'$ by $E$ in 
$\langle {\bf p}'_3|\left(\hat{G}_V(2E_i-E')\right)^\dagger|{\bf k}'_3\rangle$.

For the elastic diagram, Fig.~\ref{fig:trace1}b), we obtain:
\begin{eqnarray}
& & {\bf J}^{\rm (fig.\ref{fig:trace1}b)}({\bf r})  =  \left(\frac{1}{2}\right)^3 2^3\int\frac{{\rm d}{\bf k}{\rm d}{\bf k}_1{\rm d}{\bf k}_2{\rm d}{\bf k}_3{\rm d}{\bf k}'_1{\rm d}{\bf k}_2' {\rm d}{\bf k}'_3{\rm d}{\bf p}_1{\rm d}{\bf p}_2{\rm d}{\bf p}_3{\rm d}{\bf p}_4}{(2\pi)^{33}}
\nonumber\\
&  &\times \int_{-\infty}^\infty \frac{{\rm d}E}{2\pi i} w^*({\bf k}'_1)w^*({\bf k}'_2)w({\bf k}_1)w({\bf k}_2)\langle{\bf k}'_1
|\left(\hat{\Omega}^{(V)}(E_i)\right)^\dagger|{\bf k}'_3\rangle \langle{\bf k}'_3|\hat{\bf J}({\bf r})|{\bf k}_3\rangle
\nonumber\\
& & \times
\langle{\bf k}'_2|\left(\hat{\Omega}^{(V)}(E_i)\right)^\dagger|{\bf k}\rangle
\langle {\bf k}|\hat{G}_V(E)|{\bf p}_4\rangle
\langle {\bf k}_3|\hat{G}_V(2E_i-E)|{\bf p}_3\rangle \nonumber\\
& & \times  \langle{\bf p}_3,{\bf p}_4|\hat{T}_U(2E_i)|{\bf p}_1,{\bf p}_2\rangle 
\langle{\bf p}_1|\hat{\Omega}^{(V)}(E_i)|{\bf k}_1\rangle\langle{\bf p}_2|\hat{\Omega}^{(V)}(E_i)|{\bf k}_2\rangle  \label{eq:figtrace1b}\,.
\end{eqnarray}
Here, the labels of the wave vectors are identical to Eq.~(\ref{eq:figtrace1a}), with the only difference, that the intermediate wave vectors ${\bf p}^\prime_1,\dots,{\bf p}^\prime_4$ are not needed due to the missing interaction event for the dashed amplitudes. The trace formula, Eq.~(\ref{eq:trace4}), is applied as follows: (i) replace 
$\langle{\bf k}'_2|\left(\hat{\Omega}^{(V)}(E_i)\right)^\dagger|{\bf k}\rangle
\langle {\bf k}|\hat{G}_V(E)|{\bf p}_4\rangle$ by $\langle{\bf k}'_2|\left(\hat{\Omega}^{(V)}(E_i)\right)^\dagger|{\bf p}_4\rangle$,
(ii) delete the integrals $\int{\rm d}{\bf k}/(2\pi)^3$ and $\int{\rm d}E/(2\pi i)$, and (iii) replace $E$ by $E_i$ in $\langle {\bf k}_3|\hat{G}_V(2E_i-E)|{\bf p}_3\rangle$.

It is also insightful to take a look at the prefactors: the first factor $1/2$ in Eq.~(\ref{eq:figtrace1a}) results from $(1/\sqrt{2})^2$ in the initial states
$|i\rangle$ and $\langle i|$, see Eq.~(\ref{eq:initial}). The integration over the final states $|{\bf k}_3,{\bf k}_4\rangle$ and 
$\langle {\bf k}'_3,{\bf k}'_4|$ (with ${\bf k}_4={\bf k}'_4={\bf k}$ due to the trace) goes along with two more factors $1/2$
(since both integrations must be performed in the symmetrized subspace). This, however, is counterbalanced by the fact that we may select either
one of the two final particles as the detected particle. In Eq.~(\ref{eq:figtrace1a}), we have selected $|{\bf k}_3\rangle$ and $\langle{\bf k}'_3|$. Therefore, we have to include a factor $2^2$ to take into account the other possibilities. In Eq.~(\ref{eq:figtrace1b}), we obtain an additional factor $2$ due to the two possibilities in the factorization formula, Eq.~(\ref{eq:factorizeom}), for the dashed amplitudes: ${\bf k}'_1$ can be associated with
${\bf k}_3'$ and ${\bf k}'_2$ with ${\bf k}_4'={\bf k}$ -- or vice versa. 

Finally, the diagrams shown in Fig.~\ref{fig:trace1} can be generalized to $N>2$ particles. In this case, the remaining $N-2$ particles are assumed not to interact with the detected particle. Hence, their evolution factorizes from the one of the detected particle and need not be taken into account. The prefactors are then generalized as follows: $1/2\to N(N-1)/4$ in Eq.~(\ref{eq:figtrace1a}) and
 $1\to N(N-1)/2$  in Eq.~(\ref{eq:figtrace1b}). Let us now compare these prefactors with the ones obtained from the iterative procedure based on the connection of  building blocks in Secs.~\ref{sec:incoherent} and \ref{sec:coherent}: The factors $N(N-1)\simeq N^2$ (for $N\gg 1$, since $N\to\infty$ in the quasi-stationary limit) are accounted for by the source term $\rho_0$ in Eq.~(\ref{eq:i0}), which is proportional to $N$, see Eq.~(\ref{eq:rho0}), and occurs two times for a two-particle process  proportional to the  density squared. What remains is a factor $1/2$ for each collision event [twice in Fig.~\ref{fig:trace1}a) and once in Fig.~\ref{fig:trace1}b)] which is included in the definition of the building blocks, Eqs.~(\ref{eq:g},\ref{eq:f}).
The origin of this factor can be traced back to the indistinguishability of bosonic particles. Indeed, as argued at the end of 
Sec.~\ref{sec:nparticle}, all factors related to  indistinguishability finally drop out in the case where all particles are initially in the same state. Since the $T$-matrix for indistinguishable particles, see Eq.~(\ref{eq:tmatrix}), differs by a factor 2 from the one for distinguishable particles, this must be counterbalanced by the above factor $1/2$.

\section{Trace formulas}
\label{sec:trace}

Here, we prove the trace formulas, Eqs.~(\ref{eq:trace3},\ref{eq:trace4}), for the trace over the undetected particle originating from an inelastic or an elastic collision.
In both cases, we apply first the completeness relation $\int{\rm d}{\bf k}|{\bf k}\rangle\langle{\bf k}|=(2\pi)^3$, and then the following identity
for the product of two Green's operators:
\begin{equation}
\hat{G}_V^\dagger(E')\hat{G}_V(E)=\frac{1}{E-E'+i\epsilon}
\left(\hat{G}_V^\dagger(E')-\hat{G}_V(E)\right)\,,
\end{equation}
which is  similar to Eq.~(\ref{eq:gg}). Thereby,  Eq.~(\ref{eq:trace3}) is proven as follows:
\begin{eqnarray}
& & \int_{-\infty}^\infty \frac{{\rm d}E{\rm d}E'}{|2\pi i|^2}\int\frac{{\rm d}{\bf k}}{(2\pi)^3}\bigl(\dots\bigr)^{(l)}_{(-E')}
\hat{G}_V^\dagger(E')|{\bf k}\rangle\langle{\bf k}|\hat{G}_V(E)\bigl(\dots\bigr)^{(r)}_{(-E)}\nonumber\\
& & = \int_{-\infty}^\infty \frac{{\rm d}E{\rm d}E'}{|2\pi i|^2}\bigl(\dots\bigr)^{(l)}_{(-E')}
\hat{G}_V^\dagger(E')\hat{G}_V(E)\bigl(\dots\bigr)^{(r)}_{(-E)}\nonumber\\
& & =\int_{-\infty}^\infty \frac{{\rm d}E{\rm d}E'}{|2\pi i|^2} \frac{1}{E-E'+i\epsilon}
\bigl(\dots\bigr)^{(l)}_{(-E')}\left( \hat{G}_V^\dagger(E')-\hat{G}_V(E) \right) \bigl(\dots\bigr)^{(r)}_{(-E)}\nonumber\\
& & =
\int_{-\infty}^\infty \frac{{\rm d}E}{2\pi i} \bigl(\dots\bigr)^{(l)}_{(-E)} \left(\hat{G}_V^\dagger(E)-\hat{G}_V(E)\right)
\bigl(\dots\bigr)^{(r)}_{(-E)}\label{eq:apptrace1}\,.
\end{eqnarray}
In the last step, we have used the fact that $\bigl(\dots\bigr)^{(r)}_{(-E)}$ is a complex analytic function without poles in the lower half of the complex plane. Similarly, $\bigl(\dots\bigr)^{(l)}_{(-E')}$ exhibits no poles in the upper half. Thereby, considering the two terms $\hat{G}_V^\dagger(E')$ or $\hat{G}_V(E)$, respectively, we can perform the integral either over $E$ or over $E'$, closing the integration contour in the lower or upper half, respectively. In both cases, the term $1/(E-E'+i\epsilon)$ gives the only pole.
This fixes $E'=E$, and we arrive at the final result, Eq.~(\ref{eq:apptrace1}).

Concerning the trace formula for elastic collisions, Eq.~(\ref{eq:trace4}), we proceed in a similar way as in Eq.~(\ref{eq:omgmerge}). We  use the definition of $\hat{\Omega}_+^{(V)}(E_i)$, Eq.~(\ref{eq:omegav}), and 
the Lippmann-Schwinger equation (\ref{eq:lippmannschwingergv}) for  $\hat{G}_V(E)$ as follows:
\begin{eqnarray}
& & \int_{-\infty}^\infty \frac{{\rm d}E}{2\pi i}\int\frac{{\rm d}{\bf k}}{(2\pi)^3}
\langle{\bf k}_i|\left(\hat{\Omega}_+^{(V)}(E_i)\right)^\dagger|{\bf k}\rangle\langle{\bf k}| \hat{G}_V(E) \bigl(\dots\bigr)_{(-E)}\nonumber\\
& & = \int_{-\infty}^\infty \frac{{\rm d}E}{2\pi i} \langle{\bf k}_i| \left[\hat{G}_V(E) +\hat{V}\hat{G}_V^\dagger(E_i)\hat{G}_V(E)\right]\bigl(\dots\bigr)_{(-E)}\nonumber\\
& & =\int_{-\infty}^\infty \frac{{\rm d}E}{2\pi i} \frac{1}{E-E_i+i\epsilon} \langle{\bf k}_i|
\left[{\mathbbm 1}+\hat{V}\left(\hat{G}_V(E)+\hat{G}_V^\dagger(E_i)-\hat{G}_V(E)\right)\right] \bigl(\dots\bigr)_{(-E)}\nonumber\\
& & =\langle{\bf k}_{i}|\left[{\mathbbm 1}+\hat{V}\hat{G}_V^\dagger(E_i)\right]\bigl(\dots\bigr)_{(-E_i)}=
\langle{\bf k}_{i}|\left(\hat{\Omega}_+^{(V)}(E_i)\right)^\dagger\bigl(\dots\bigr)_{-(E_i)}\,.
\end{eqnarray}
This proves Eq.~(\ref{eq:trace4}).

\section{Particle and energy flux conservation}
\label{sec:cons}

In this appendix, we prove Eqs.~(\ref{eq:particlecons}) and (\ref{eq:energycons}). Starting from Eq.~(\ref{eq:fsimple}) for
$f_{E_1,E_2,E_3}$, we calculate $\int_{0}^\infty {\rm d}E_3\sqrt{E_3}f_{E_1,E_2,E_3}$. For this purpose, we first note that:
\begin{eqnarray}
& & \int_0^\infty{\rm d}E_3 \sqrt{E_3}\left( \frac{G^*_{E_1+E_2-E_3}(k_4)-G_{E_1+E_2-E_3}(k_4)}{2\pi i}\right)\left|G_{E_3}(k_3)\right|^2\nonumber\\
&   & = \int_{-\infty}^\infty {\rm d}E_3 \left( \frac{G^*_{E_1+E_2-E_3}(k_4)-G_{E_1+E_2-E_3}(k_4)}{2\pi i}\right)
\left(\frac{G^*_{E_3}(k_3)-G_{E_3}(k_3)}{2i/\ell_{\rm dis}}\right)\nonumber\\
&&\simeq \frac{\ell_{\rm dis}}{2i}\left( \frac{1}{E_1+E_2-k_3^2-k_4^2-2i\varepsilon}-\frac{1}{E_1+E_2-k_3^2-k_4^2+2i\varepsilon}\right) \nonumber\\ 
& & \simeq \frac{\ell_{\rm dis}}{2i}\left(\left[G^{(0,m/2)}_{E_{12}}(({\bf k}_3-{\bf k}_4)/2)\right]^*-G^{(0,m/2)}_{E_{12}}(({\bf k}_3-{\bf k}_4)/2)\right)
\label{eq:appb2}\,,
\end{eqnarray}
with $E_{12}=E_1+E_2-E_{{\bf k}_1+{\bf k}_2}/2$, ${\bf k}_1+{\bf k}_2={\bf k}_3+{\bf k}_4$ and $|{\bf k}_{34}\rangle$ as defined after Eq.~(\ref{eq:tmatrix}).
Here, we have first used the identity $\sqrt{E_3}|G_{E_3}(k_3)|^2=\ell_{\rm dis}[G^*_{E_3}(k_3)-G_{E_3}(k_3)]/(2i)$ for the average Green's function, then 
replaced the average Green's functions by vacuum Green's functions (which is appropriate in the weak disorder limit), and evaluated the integral over $E_3$ using residual calculus. By setting the imaginary part $2\varepsilon$ in the denominator to $\varepsilon$ and using momentum conservation, i.e.~$(k_3^2+k_4^2)/2\to E_{{\bf k}_1+{\bf k}_2}/2-{\bf k}_3{\bf k}_4$ we arrived at Eq.~(\ref{eq:appb2}). Again, $G_E^{(0,m/2)}({\bf k})=1/(E-2k^2+i\epsilon)$ denotes the vacuum Green's function for a particle with mass $m/2$, cf.~Eq.~(\ref{eq:opttheorem}).

Inserting Eq.~(\ref{eq:appb2}) into Eq.~(\ref{eq:fsimple}), and substituting the variable ${\bf k}_3\to {\bf k}_{34}=({\bf k}_3-{\bf k}_4)/2$, the integration over 
${\bf k}_{34}$ reduces to: 
\begin{eqnarray}
& & 2\int\frac{{\rm d}{\bf k}_{34}}{(2\pi)^{3}}
\langle {\bf k}_{34}|\left(\hat{G}_{0,m/2}^\dagger(E_{12})-\hat{G}_{0,m/2}(E_{12})\right)|{\bf k}_{34}\rangle 
\left|\langle{\bf k}_{34}|\hat{T}^{(1)}_U(E_{12})|{\bf k}_{12}\rangle\right|^2
\nonumber\\
& & =2 \langle{\bf k}_{12}|\left(\hat{T}^{(1)}_U(E_{12})\right)^\dagger\left(\hat{G}_{0,m/2}^\dagger(E_{12})-\hat{G}_{0,m/2}(E_{12})\right)\hat{T}^{(1)}_U(E_{12})|{\bf k}_{12}\rangle\label{eq:appb4}\,.
\end{eqnarray}
Applying the
optical theorem, Eq.~(\ref{eq:opttheorem}), yields in total:
\begin{eqnarray}
\int_0^\infty {\rm d}E_3\sqrt{E_3}f_{E_1,E_2,E_3}  & = &   - \frac{(4\pi)^2}{\ell_{\rm dis}}
\int\frac{{\rm d}{\bf k}_1{\rm d}{\bf k}_2}{(2\pi)^{6}} ~{\rm Im}\left\{\langle{\bf k}_{12}|\hat{T}^{(1)}_U(E_{12})|{\bf k}_{12}\rangle\right\}
\nonumber\\
& & \times 
\left|G_{E_1}(k_1)\right|^2 \left|G_{E_2}(k_2)\right|^2\label{eq:appb}\,.
\end{eqnarray}
On the other hand, the integral over ${\bf r}_1$ and ${\bf r}_2$ in Eqs.~(\ref{eq:g},\ref{eq:gcontact}), see Eq.~(\ref{eq:gsimple}), together with the formula
$\sqrt{E_2}\left|G_{E_2}(k_2)\right|^2= \ell_{\rm dis}[G^*_{E_2}(k_2)-G_{E_2}(k_2)]/(2i)$, yields:
\begin{eqnarray}
-\sqrt{E_2}g_{E_1,E_2}& = &- \frac{(4\pi)^2}{\ell_{\rm dis}}
\int\frac{{\rm d}{\bf k}_1{\rm d}{\bf k}_2}{(2\pi)^{6}} \left|G_{E_1}(k_1)\right|^2{\rm Im}\left\{\langle{\bf k}_{12}|\hat{T}^{(1)}_U(E_{12})|{\bf k}_{12}\rangle \right.\nonumber\\
& & \times\left.\left(G^*_{E_2}(k_2)-G_{E_2}(k_2)\right)G_{E_2}(k_2) \right\}\label{eq:appb5}\,.
\end{eqnarray}
The term $|G_{E_2}(k_2)|^2$ in the second line of Eq.~(\ref{eq:appb5}) exactly reproduces Eq.~(\ref{eq:appb}), whereas
the remaining term $[G_{E_2}(k_2)]^2$ gives a negligible contribution in the limit $\sqrt{E_2}\ell_{\rm dis}\gg 1$. Moreover, one can show that this contribution is cancelled by another diagram where an additional disorder correlation function is inserted just before and after the collision, see Fig. 1(e,f) in \cite{cherroret11}. This proves Eq.~(\ref{eq:particlecons}).

Eq.~(\ref{eq:energycons}) can be shown in almost the same way. When calculating $\int_{0}^\infty {\rm d}E_3~2\sqrt{E_3}E_3f_{E_1,E_2,E_3}$, 
Eq.~(\ref{eq:appb2}) is replaced by:
\begin{eqnarray}
& & \int_0^\infty{\rm d}E_3~2\sqrt{E_3}E_3\left( \frac{G^*_{E_1+E_2-E_3}(k_4)-G_{E_1+E_2-E_3}(k_4)}{2\pi i}\right)\left|G_{E_3}(k_3)\right|^2\nonumber\\
& & \simeq \frac{\ell_{\rm dis}}{2i}\left(\left[G^{(0,m/2)}_{E_{12}}({\bf k}_{34})\right]^*-G^{(0,m/2)}_{E_{12}}({\bf k}_{34})\right)\Bigl(E_1+E_2+k_3^2-k_4^2\Bigr)\,.
\end{eqnarray}
When integrating over ${\bf k}_{34}$ as in Eq.~(\ref{eq:appb4}), the  term $(k_3^2-k_4^2)$ vanishes due to symmetry, but the factor
$(E_1+E_2)$ remains. This proves Eq.~(\ref{eq:energycons}).

\section{Collision terms for  $s$-wave scattering}
\label{sec:swave}

For a short-range interaction potential $U({\bf r})$, the $T$-matrix (for a particle with mass $m/2$) has the following form \cite{rossum99}:
\begin{equation}
\langle {\bf k}'|\hat{T}^{(1)}_U(E)|{\bf k}\rangle=8\pi a_s\left(1-i \sqrt{\frac{E}{2}} a_s + \mathcal{O}\left(\sqrt{E}a_s\right)^2\right)\,,
\label{eq:swave}
\end{equation}
valid for arbitrary plane wave states $|{\bf k}\rangle,|{\bf k}'\rangle$ in the limit $\sqrt{E}a_s\ll 1$, where $a_s$ is the $s$-wave scattering length associated to the potential $U({\bf r})$. The prefactor $8\pi a_s\equiv 4\pi a_s \hbar^2/m$ for $\hbar^2/(2m)\equiv 1$ 
applies for a particle with mass $m/2$.
 Note that, due to the symmetrization of the states
$|{\bf k}_{12}\rangle$ and $|{\bf k}_{34}\rangle$, an additional factor 2 appears when evaluating $\langle{\bf k}_{34}|\hat{T}_U^{(1)}(E_{12})|{\bf k}_{12}\rangle$  in Eq.~(\ref{eq:tmatrix}), cf.~the remark at the end of \ref{sec:examplediag}. Inserting Eq.~(\ref{eq:swave}) into the general expressions, Eqs.~(\ref{eq:g},\ref{eq:f},\ref{eq:gc},\ref{eq:fc},\ref{eq:hc}), yields the following results for the collision terms (in the limit $\ell_{\rm dis}\sqrt{E_{1,2,3}}\gg 1$):
\begin{equation}
g_{E_1,E_2} = -\frac{\alpha}{6\rho_0\sqrt{E_1}E_2}
\left[\left(\sqrt{E_1}+\sqrt{E_2}\right)^3-\left|\sqrt{E_1}-\sqrt{E_2}\right|^3\right]\label{eq:gswave}\,,
\end{equation}
\begin{equation}
f_{E_1,E_2,E_3}=\frac{\alpha}{\rho_0\sqrt{E_1E_2E_3}}
\min\left(\sqrt{E_1},\sqrt{E_2},\sqrt{E_3},\sqrt{E_1+E_2-E_3}\right)\label{eq:fswave}\,,
\end{equation}
\begin{eqnarray}
g^{(C)}_{E_1,E_2} & = & -\frac{2}{\rho_0\sqrt{E_2}}{\rm Re}\left\{ \frac{1}{\left[1-i\ell_{\rm dis}\left(\sqrt{E_2}-\sqrt{\widetilde{E}_2}\right)\right]^2}\right.\nonumber\\
& & \times\left.\left[i\beta \sqrt{E}_i +\alpha \frac{\left(\sqrt{E_1}+\sqrt{E_2}\right)^3-\left|\sqrt{E_1}-\sqrt{E_2}\right|^3}{12\sqrt{E_1E_2}}\right]\right\}
\label{eq:gcswave}\,,
\end{eqnarray}
\begin{equation}
h^{(C)}_{E_1,E_2}  =  \frac{-2i\beta\sqrt{E_i}-\alpha \sqrt{2E_i+2E_d}}{\rho_0\left[1-i\ell_{\rm dis}\left(\sqrt{E_2}-\sqrt{\widetilde{E}_2}\right)\right]\left[\sqrt{E_1}+\sqrt{\widetilde{E}_1}-i\ell_{\rm dis}(E_1-\widetilde{E}_1)\right]}\label{eq:hcswave}\,,
\end{equation}
\begin{eqnarray}
f^{(C)}_{E_1,E_2,E_3} & = & \frac{\alpha}{\rho_0\sqrt{E_1}\left[\sqrt{E_2}+\sqrt{\widetilde{E}_2}-i\ell_{\rm dis}(E_2-\widetilde{E}_2)\right]\left[\sqrt{E_3}+\sqrt{\tilde{E_3}}-i\ell_{\rm dis}(E_3-\widetilde{E}_3)\right]}\nonumber\\
& & \times \sum_{s_i\in\{0,1\}}(-1)^{s_1+s_2+s_3+s_4+1}
\left(|k_s|+\frac{2i k_s}{\pi} \ln |k_s|\right)\label{eq:fcswave}\,,
\end{eqnarray}
where
\begin{eqnarray}
k_s & = & (-1)^{s_1}\sqrt{E_1}+ (-1)^{s_2}\left[s_2\sqrt{E_2}+(1-s_2)\sqrt{\widetilde{E}_2}\right]\nonumber\\
& & +(-1)^{s_3}\left[s_3\sqrt{E_3}+(1-s_3)\sqrt{\widetilde{E}_3}\right]+(-1)^{s_4}\sqrt{E_1+E_2-E_3}\label{eq:ks}\,.
\end{eqnarray}
In the above expressions, Eqs.~(\ref{eq:gswave}-\ref{eq:ks}), all energies appearing under a square root must be positive -- otherwise, the corresponding expression is set to zero (e.g. $f_{E_1,E_2,E_3}=0$ if $E_3>E_1+E_2$). Furthermore, note that the density $\rho_0$ appearing in the denominators of Eqs.~(\ref{eq:gswave}-\ref{eq:fcswave}) drops out when expressing the densities $I_E({\bf r})$, $C^{(1)}_E({\bf r})$ and $C^{(2)}_E({\bf r})$ in Eqs.~(\ref{eq:transport2},\ref{eq:c1},\ref{eq:c2})
 in units of the incoming density $\rho_0$, see also Eqs.~(\ref{eq:jsce},\ref{eq:jscce}). Therefore, the effective strength of the collision terms, Eqs.~(\ref{eq:gswave}-\ref{eq:fcswave}), is solely governed by the parameters $\alpha$ and $\beta$ introduced in Eqs.~(\ref{eq:alpha},\ref{eq:beta}).


\section*{References}
\begin{thebibliography}{10}

\bibitem{clement05}Cl\'ement D, Var\'on A F, Hugbart M,  Retter J A, Bouyer P, Sanchez-Palencia L, Gangardt D M, Shlyapnikov G V, and Aspect A 2005 {\it Phys. Rev. Lett.} {\bf 95} 170409

\bibitem{fort05}Fort C, Fallani L, Guarrera V, Lye J E, Modugno M, Wiersma D S, and Inguscio M 2005, {\it Phys. Rev. Lett.} {\bf 95} 170410

\bibitem{schulte05}Schulte T, Drenkelforth S, Kruse J, Ertmer W, Arlt J, Sacha K, Zakrzewski J, and Lewenstein M 2005 {\it Phys. Rev. Lett.} {\bf 95} 170411

\bibitem{anderson58}Anderson P W 1958 {\it Phys. Rev.} {\bf 109} 1492

\bibitem{jendrzejewski11}Jendrzejewski F, Bernard A, M\"uller K, Cheinet P, Josse V, Piraud M, Pezze L, Sanchez-Palencia L, Aspect A, and Bouyer P 2011 {\it Nat. Phys.} {\bf 8} 398

\bibitem{kondov11}Kondov S S, McGehee W R, Zirbel J J, and DeMarco B 2011 {\it Science} {\bf 334} 66

\bibitem{bergmann84}Bergmann G 1984 {\it Phys. Rep.} {\bf 107} 1

\bibitem{kuga84}Kuga Y and Ishimaru A 1984 {\it J. Opt. Soc. Am. A} {\bf 1} 831

\bibitem{albada85}van Albada M P and Lagendijk A 1985 {\it Phys. Rev. Lett.} {\bf 55} 2692

\bibitem{wolf85}Wolf P-E and Maret G 1985 {\it Phys. Rev. Lett.} {\bf 55} 2696

\bibitem{labeyrie12}Labeyrie G, Karpiuk T, Schaff J-F, Gr\'emaud B, Miniatura C, and Delande D 2012 {\it Europhys. Lett.} {\bf 100} 66001

\bibitem{jendrzejewski12}Jendrzejewski F, M\"uller K, Richard J, Date A, Plisson T, Bouyer P, Aspect A, and Josse V 2012 {\it Phys. Rev. Lett.} {\bf 109} 195302

\bibitem{karpiuk12}Karpiuk T, Cherroret N, Lee K L, Gr\'emaud B, M\"uller C A, and Miniatura C
 2012 {\it Phys. Rev. Lett.} {\bf 109} 190601
 
\bibitem{shepelyansky94}Shepelyansky D L 1994 {\it Phys. Rev. Lett.} {\bf 73} 2607
 
\bibitem{ivanchenko11}Ivanchenko M V, Laptyeva T V, and Flach S 2011 {\it Phys. Rev. Lett.} {\bf 107} 240602

\bibitem{lee90}Lee D K K and Gunn J M F 1990 {\it J. Phys. Cond. Mat.} {\bf 2} 7753

\bibitem{huang92}Huang K and Meng H-F 1992 {\it Phys. Rev. Lett.} {\bf 69} 644

\bibitem{giorgini94}Giorgini S, Pitaevskii L, and Stringari S 1994 {\it Phys. Rev. B} {\bf 49} 12938

\bibitem{bilas06}Bilas N and Pavloff N 2006 {\it Eur. Phys. J. D} {\bf 40} 387

\bibitem{gaul11}Gaul C. and M\"uller C A 2011 {\it Phys. Rev. A} {\bf  83} 063629

\bibitem{zagrebnov01}Zagrebnov V A and Bru J-B 2001 {\it Phys. Rep.} {\bf  350} 291

\bibitem{guerin06}Guerin W, Riou J-F, Gaebler J P, Josse V, Bouyer P, and Aspect A 2006 {\it Phys. Rev. Lett.} {\bf 97} 200402

\bibitem{couvert08}Couvert A, Jeppesen M, Kawalec T, Reinaudi G, Mathevet R, and Gu\'ery-Odelin D 2008 {\it Europhys. Lett.} {\bf 83} 50001

\bibitem{kuhn05}Kuhn R C, Miniatura C, Delande D, Sigwarth O, and M\"uller C A 2005 {\it Phys. Rev. Lett.} {\bf 95} 250403

\bibitem{kuhn07}Kuhn R C, Sigwarth O, Miniatura C, Delande D, and M\"uller C A 2007 {\it New J. Phys.} {\bf 9} 161

\bibitem{paul05}Paul T, Leboeuf P, Pavloff N, Richter K, and Schlagheck P 2005 {\it Phys. Rev. A} {\bf 72} 063621

\bibitem{hartung08}Hartung M, Wellens T, M\"uller C A, Richter K, and Schlagheck P 2008 {\it Phys. Rev. Lett.} {\bf 101} 020603

\bibitem{wellens09b}Wellens T and Gr\'emaud B 2009 {\it Phys. Rev. A} {\bf 80} 063827

\bibitem{wellens09c}Wellens T  2009 {\it Appl. Phys. Lett.} {\bf 95} 189

\bibitem{ernst10}Ernst T, Paul T, and Schlagheck P 2010 {\it Phys. Rev. A} {\bf 81} 013631

\bibitem{geiger12}Geiger T, Wellens T, and Buchleitner A 2012 {\it Phys. Rev. Lett.} {\bf 109} 030601

\bibitem{uehling33}Uehling E A and Uhlenbeck G E 1933 {\it  Phys. Rev.} {\bf 43} 552

\bibitem{spohn07}Spohn H 2007 arXiv:0706.0807.

\bibitem{benedetto08} Benedetto D, Castella F, Esposito R, and Pulvirenti M 2008 {\it Comm. Math. Phys.} {\bf 277} 1

\bibitem{kirkpatrick83}Kirkpatrick T R and Dorfman J R 1983 {\it Phys. Rev. A} {\bf 28} 2576

\bibitem{zaremba99}Zaremba E, Nikuni T, and Griffin A 1999 {\it J. Low Temp. Phys.} {\bf 116} 277

\bibitem{gardiner97}Gardiner C W and Zoller P 1997 {\it  Phys. Rev. A} {\bf 55} 2902

\bibitem{walser99}Walser R, Williams J, Cooper J, and Holland M 1999 {\it Phys. Rev. A} {\bf 59} 3878

\bibitem{schelle11}Schelle A, Wellens T, Delande D, and Buchleitner A 2011 {\it Phys. Rev. A} {\bf  83} 013615

\bibitem{proukakis01}Proukakis N P 2001 {\it J. Phys. B} {\bf  34} 4737

\bibitem{wachter01}Wachter J, Walser R, Cooper J, and Holland M 2001 {\it Phys. Rev. A} {\bf  64} 053612

\bibitem{erdoes12} Erd\"os L 2012 {\em Lecture notes on quantum Brownian motion}, in {\em Quantum Theory from Small to Large Scales}, edited by Fr\"ohlich J,  Salmhofer M, Mastropietro V, de~Roeck W, and Cugliandolo L F (Oxford University Press, Oxford, UK)

\bibitem{akkermans07} Akkermans E and Montambaux G  2007 {\em Mesoscopic Physics of Electrons and Photons} (Cambridge University Press, Cambridge, UK)

\bibitem{taylor}Taylor J R 1972 {\em Scattering Theory: The Quantum Theory on Nonrelativistic Collisions} (John Wiley \& Sons, New York)

\bibitem{ishimaru}Ishimaru A 1978 {\em Wave Propagation and Scattering in Random Media} (Academic, New York) Vols. I and II.

\bibitem{tichy10}Tichy M C, Tiersch M, de Melo F, Mintert F, and Buchleitner A 2010 {\it Phys. Rev. Lett} {\bf 104} 220405

\bibitem{spectral}Skipetrov S E, Minguzzi A, van Tiggelen B A, and Shapiro B 2008 {\it Phys. Rev. Lett} {\bf 100} 165301

\bibitem{rammer}Rammer 2004 {\it Quantum Transport Theory} (Westview Press)

\bibitem{rossum99}van Rossum M C W and Nieuwenhuizen T M 1999 {\it Rev. Mod. Phys.} {\bf 71} 313

\bibitem{recurrent}Wiersma D S, van Albada M P, van Tiggelen B A, and Lagendijk A 1995 {\it Phys. Rev. Lett.} {\bf 74} 4193 

\bibitem{schwiete13b}Schwiete G and Finkel'stein A M, arXiv:1302.0028

\bibitem{langer66}Langer J S and Neal T 1966 {\it Phys. Rev. Lett.} {\bf 16} 984

\bibitem{akkermans}Akkermans E , Wolf P E, and Maynard R 1986 {\it Phys. Rev. Lett.} {\bf 56} 1471

\bibitem{mark88}van der Mark M B, van Albada M P, and Lagendijk A 1988 {\it Phys. Rev. B} {\bf 37} 3575

\bibitem{pethick08} Pethick C J and Smith H 2008 {\em Bose-Einstein condensation in dilute gases} (Cambridge University Press, Cambridge, UK)

\bibitem{lieb05}Lieb E H, Seiringer R, Solovej J P, and Yngvason J 2005 {\em The mathematics of the Bose gas and its condensation} (Birkh\"auser, Basel)

\bibitem{wellens06}Wellens T, Gr\'emaud B, Delande D, and Miniatura C 2006 {\it Phys. Rev. A} {\bf 73} 013802

\bibitem{wellens06b}Wellens T and Gr\'emaud B 2006 {\it J. Phys. B} {\bf 39} 4719

\bibitem{schwiete13}Schwiete G and Finkel'stein A M, arXiv:1301.1925

\bibitem{fortagh07}Fort\'agh J and Zimmermann C 2007 {\it Rev. Mod. Phys.} {\bf 79} 235

\bibitem{schwiete10}Schwiete G and Finkel'stein A M 2010 {\it Phys. Rev. Lett.} {\bf 104} 103904 

\bibitem{cherroret11}Cherroret N and Wellens T 2011 {\it Phys. Rev. A} {\bf 84} 021114


\end{thebibliography}
\end{document}